\documentclass[preprint,aps,nofootinbib]{revtex4-2}
\pdfoutput=1
\usepackage{graphicx}% Include figure files
\usepackage{epstopdf}    % nado dobavit, inache pdf ne budet
\usepackage{dcolumn}% Align table columns on decimal point
\usepackage{amsfonts}
\usepackage{amsmath}
\usepackage{amssymb}
\usepackage{graphicx}
\usepackage{bm}% bold math
\usepackage{color}
\usepackage{comment}
\usepackage{float}
\usepackage[caption=false]{subfig}
\usepackage{enumitem}
\usepackage{lipsum} % just for the example
\usepackage{epsf}
\usepackage{float}
\usepackage{enumitem}
\usepackage[utf8]{inputenc}

\newcolumntype{L}{>{\centering\arraybackslash}m{3cm}}

\newcommand{\edc}{\end{document}}
\newcommand{\bb} {\begin{thebibliography}}
\newcommand{\eb}{\end{thebibliography}}
\newcommand{\bc}{\begin{center}}
\newcommand{\ec}{\end{center}}

\newcommand{\be}{\begin{equation}\small}
\newcommand{\ee}{\end{equation}\normalsize}
\newcommand{\bea}{\begin{eqnarray}}
\newcommand{\eea}{\end{eqnarray}}
\newcommand{\ba}{\begin{array}{l}   }
\newcommand{\lab}[1]{\label{#1}}
\newcommand{\ea}{\end{array}}
\newcommand{\dsfrac}{\displaystyle\frac}
\newcommand{\ds} {\displaystyle}
\newcommand{\summa}{\ds\sum}
\newcommand{\dssum}{\summa}
\newcommand{\re}[1]{(\ref{#1})}

\newcommand{\ci}{\cite}

\newcommand{\dsint}{\ds\int}

\def\bfr{{\bf r}}
\def\bfk{{\bf k}}

\newcommand{{\vergul}}{  ,}

\newcommand{\veps}{\varepsilon }

\newcommand{\cale}{{\cal E}}

\begin{document}
	%\twocolumn
	%\sloppy
	\draft
	%\doublespace
	\title{Resolving the problem of complex sound velocity
		in binary Bose mixtures with attractive
		intercomponent interactions}
	\author{Abdulla Rakhimov$^{1,2}$, Sanathon Tukhtasinova$^{3}$,  and Vyacheslav I. Yukalov$^{4,5}$}
	
	%\email{rakhimovabd@yandex.ru,  abdurakhmonov.t.z@gmail.com,  yukalov@theor.jinr.ru}
	
	%\address{
		\affiliation{
			$^1$Institute of Nuclear Physics,Academy of Science of Uzbekistan, Tashkent 100214, Uzbekistan \\
			$^2$ Center for Theoretical Physics, Khazar University, 41 Mehseti 
			Street, Baku, AZ1096, Azerbaijan\\
			$^3$Physics Department, National University of Uzbekistan, Tashkent, 100084, Uzbekistan \\
			$^4$Bogoliubov Laboratory of Theoretical Physics, Joint Institute for Nuclear Research, 
			Dubna 141980, Russia \\
			$^5$Instituto de Fisica de S\~ao Carlos, Universidade de S\~ao Paulo, \\
			CP 369, S\~ao Carlos 13560-970, S\~ao Paulo, Brazil}

		\date{\today}

%%%%%%%%%%%%%%%%%%%%%%%%%%%%%%%%%
\begin{abstract}

In 2015  Dmitry  Petrov theoretically suggested that, in a binary mixture of bosons
a quantum liquid droplet may arise due to the competition between attractive intercomponent
and repulsive intracomponent forces. Although this prediction has been confirmed
experimentally, the model by itself suffers from a serious conceptual problem:
The low - lying excitation spectrum manifests  a purely imaginary phonon velocity,
$c_d^2 < 0$. In the present work, we develop a self consistent theory of two-component Bose
systems with attractive interspecies interactions, which accurately takes into account pair
correlations in terms of anomalous and mixed densities. We have shown that this procedure is
able to resolve the problem of  $c_d^2 < 0$. Limiting ourselves with a symmetric Bose mixture
at zero temperature, we have found a region of stability in which a droplet can survive.

\end{abstract}

\maketitle

\section{Introduction}
\label{sec1}

More than hundred years ago Bose and Einstein predicted that, particles with integer spins
in an ideal Bose gas could occupy the ground state macroscopically. This phenomenon, which
later became known as Bose -- Einstein condensation (BEC), marked the beginning for a
deeper understanding of the nature of quantum liquids, such as superfluidity and
superconductivity. It has been shown \cite{Bog47} that the presence of interactions
strongly influences the properties of BEC, modifying the energy dispersion from the simple
quadratic expression $\varepsilon_k={\bf{k}}^2/{2m}$ to the spectrum that is linear in the
long-wave limit \footnote{Here and below we set $\hbar=1$ and $k_B=1$.}, having in the
Bogolubov approximation the form
\begin{equation}
    E_k \approx \sqrt{\varepsilon_k(\varepsilon_k + 2g\rho)} \; ,
    \label{EBog}
\end{equation}
with the sound velocity
\begin{equation}
    c=\sqrt{2g\rho/m}
    \label{CBog} \; .
\end{equation}
In Eqs. \re{EBog} and \re{CBog}, $m$ is the particle mass, $\rho$, particle density
$\rho=N/V$, and the contact interaction $V({\bf{r}}-{\bf{r}}') = g\delta ({\bf{r}}-{\bf{r}}')$
is kept in mind, with  the coupling constant of the contact interaction $g = 4\pi a_s/m$.
Particularly, it has been observed that, when the $s$ -- wave scattering length is suddenly
changed into a negative value, the BEC collapses and undergoes an explosion in which a
substantial fraction of atoms are blown off (Bosenova phenomenon) \cite{Donley}. Thus,
for many years, the dominant idea has been that attractive forces have only destructive
influence on BEC.

However, Petrov \ci{petrov} noticed that, in a two component mixture of BEC the attraction
between different atoms may lead to the formation of a new state of matter -- quantum
liquid droplets (droplets for further reference). In fact, taking into account quantum
fluctuations resulting in the Lee -- Huang -- Yang (LHY) energy terms, he predicted that,
in a two -- component Bose system a stable droplet may arise due to the balance between
the attractive intercomponent (${g_{ab}}<0$) and repulsive intracomponent
(${g_{aa}} >0, {g_{bb}>0}$) interactions. This occurs when the attractive ground-state term
$\cale_{GS}<0$ is compensated by the repulsive LHY energy $\cale_{LHY}>0$. Soon after the
theoretical prediction, quantum liquid droplets in two component Bose mixtures
were observed experimentally \cite{semegini,cabrera,errico,burchanti,guo}.

Despite its important predictive power, Petrov's model has a serious shortcoming: In the
regime of droplet formation, the energy functional becomes complex due to the emergence
of a purely imaginary phonon velocity. In fact, within the Bogolubov theory, the
long -- wavelength modes of the excitation spectrum, describing the total and relative
density fluctuations, are given by the linear phonons with the speeds of sound
\begin{equation}
\ba
%    \begin{array}{cc}
           c_d=\sqrt{\dsfrac{\rho \delta g}{2m}}  , \quad\quad
           c_s=\dsfrac{\sqrt{(2-\alpha)\rho g}}{\sqrt{2m}} \; ,
         \label{Cd Cs petr}
 \ea
\end{equation}
where, for simplicity of illustration, the symmetric mixture is considered, with
$g_{aa}=g_{bb}=g$, $\rho_a=\rho_b=\rho/2$, and the standard notations for
$\delta g=g_{ab}+g$ and ${\alpha}=\delta g/g$ are introduced. It is seen that, for
$\delta g<0$ the speed of sound for the density mode $c_d$ becomes purely imaginary.
This failure has caused a lot of challenges in the literature, raising several attempts
where the authors proposed different solutions. Below, we outline some of them, explaining
why these attempts are not completely successful.

\begin{enumerate}
\item[1)] First of all, let us note that, Petrov limited himself to neglecting $c_d$ in
the $\cale_{LHY}$ term,
 \begin{equation}
     \cale_{LHY}=\dsfrac{8 m^4}{15\pi^2}({c_d}^5+{c_s}^5)\longrightarrow \dsfrac{8 m^4}{15\pi^2} {c_s}^5 \; .
     \label{Elhypetr}
 \end{equation}
Strictly speaking, this makes the consideration not self-consistent. Although the basic
Petrov's idea of taking account of fluctuations is correct, however not all important
fluctuations are taken into account in the LHY approximation. For instance, the paring
fluctuations, characterized by the so-called anomalous averages, are disregarded in the
LHY approximation, while they can be of the same order or even larger.

\item[2)]
  Ota and Astrakharchik \cite{astra} proposed to redefine the sound velocities by using
the relation between the latter and  the compressibilities:
 \begin{equation}
%     \begin{array}{cc}
 \ba
           c_{\pm}=(m\rho \kappa_{\pm})^{-\frac{1}{2}}, \quad \quad
               \kappa_\pm=\dsfrac{1}{\rho^2}\left[\dsfrac{\partial^2
               	\cale/V}{\partial(\rho_a\pm \rho_b)^2}\right]^{-1} \; ,
          \label{Chi}
     \end{array}
 \end{equation}
where
\begin{equation}
    \dsfrac{\cale}{V}=\dsfrac{g}{2}
     (\rho_a^2+\rho_b^2)+g_{ab}\rho_a\rho_b+\frac{\cale_{LHY}}{V} \; .
    \label{Etot Astra}
\end{equation}
Then the redefined density sound velocity is given by $c_d=c_+=\sqrt{c_d^2}$, with
\begin{equation}
    c_d^2=\dsfrac{\rho g}{2m}\left[\alpha+\dsfrac{4\sqrt{2\gamma}(2-\alpha)^{5/2}}{\sqrt{\pi}}
    \right] \; ,
    \label{dAstra}
\end{equation}
where $\gamma=\rho a_s^3$. It is seen that, $c_d^2\ge0$ for
\begin{equation}
    \gamma_{crit}\ge \frac{\alpha^2 \pi}{32(2+|\alpha|)^5} \; ,
\end{equation}
even for $\alpha=\delta g/g<0$.

However, this is clearly a phenomenological trick that is not well justified, since,
strictly speaking, two types of susceptibilities cannot be defined for a system. Moreover,
the inclusion of higher-order terms into the energy functional would lead to different
redefined expressions for the sound velocities. Even more, this does not remove the instability,
since the speed of sound remains imaginary in the Bogolubov approximation. The problem is again
the same--the pairing fluctuations are not included.

\item[3)]
 Gu and Yin \cite{Guin} considered the binary mixture in the regime of asymptotically
weak interactions using the Beliaev theory \cite{belyaev}, taking into account the
interactions between phonon (density) and pseudo-spin excitations. They showed that
the corrections to the phonon spectrum, coming from the interaction between phonon and
pseudo-spin excitations, can stabilize the phonon mode for a specially chosen system
density. However, since they consider a uniform system, where particle density is a
free parameter, the existence of stability solely for a particular chosen density is
unphysical.

\item[4)]
Hu and Liu \cite{Hulu} considered the bosonic pairing theory, assuming that the main
contribution comes from the inter-species anomalous averages. Note that, long before,
for a single-component condensed system, the similar approximation was employed by Girardeau
and Arnowitt \cite{Girardeau} who have got a gap in the spectrum, which contradicts the
Hugenholtz-Pines relations \cite{HP} for binary mixtures \cite{Watabe,Nepom}. A similar gap also
arises in the paper by Hu and Liu. In fact, the authors applied the Hubbard -- Stratanovich
transformation to the two-component Bose mixture by introducing an auxiliary scalar field $\Delta$.
Then, for a symmetric case, they obtained the following spectrum of quasiparticles:
\begin{equation}
    \begin{array}{cc}
         &  E_-(k)=\sqrt{\varepsilon _k(\varepsilon_k+2\mu +2\bar{\Delta})}, \\
         &  E_+(k)=\sqrt{(\varepsilon_k+2\bar{\Delta})(\varepsilon_k+2\mu +2\bar{\Delta})} \; ,
         \label{EPair}
    \end{array}
\end{equation}
where $\bar{\Delta}$ is the mean -- value of the scalar field. It is seen that, in this
bosonic pairing theory the lower branch $E_-(k)$ is gapless, while the upper branch shows
an energy gap: $E_{gap}=E_+(k=0)=2\bar{\Delta}\sqrt{1+\mu/\bar{\Delta}}$. Hence, the unstable
branch in the Petrov's theory is removed with the introduction of bosonic pairing, although a gap
appears in the other branch. However the spectrum with a gap, actually, means the condensate
instability. In this way, the Hugenholtz–Pines theorem \cite{HP,Watabe,Nepom} is not taken
into account, which implies that the chemical potential is not accurately defined, so that
the system remains unstable.

\item[5)]
Zin et al. \cite{Zin_2021} considered the stability conditions for a two-component Bose mixture,
not specifying any particular energy density functional. However, this functional is assumed
to depend only on the densities of the components, not taking account of anomalous averages
and other higher-order correlations. The studied setup is valid only under asymptotically weak
interactions and low temperature. The collapse instability is supposed to be weak, so that
a small LHY term could balance it. However, in general, the instability remains, since not all
correlations are taken into account.

\item[6)]
In the paper \cite{Zin_2022}, the authors studied a strongly dilute Bose-Bose mixture at
zero temperature and weak interactions when the terms higher in order than LHY corrections
are neglected. The instability is avoided by rejecting soft modes. The principal point of
this paper is that the authors use different forms of the chemical potential, one in the
condensate-function equation describing the condensate and another in the Bogolubov equations
defining the spectrum of collective excitations, so that to make the spectrum gapless. Using
two different chemical potentials in two different places is identical to the use of two
different chemical potentials from the beginning, as in our approach, which we specify below.
Moreover, the instability remains, as far as the higher-order correlations above LHY are
neglected.
\end{enumerate}

It can be mentioned that an attempt to take account of higher-order correlations above LHY
corrections has been undertaken by Aybar et al. \cite{Aybar_2019,Ozturk_2020}.
However, they considered a single-component Bose system. They used the Hartree-Fock-Bogolubov
approximation (HFB) in the condensate-function equation. However in the Hamiltonian the terms
of order higher then second in the operators of non-condensed atoms are omitted, thus
neglecting the anomalous average and other higher-order correlations, which corresponds to
the bilinear approximation in the calculation of the spectrum. This approximation is valid
for very weak interactions and very low temperature, when the number of non-condensed atoms
is much smaller than the total number of atoms. Unfortunately, taking different approximations
for the condensate function and for the spectrum is not self-consistent. The problem of
stability for Bose-Bose mixture was not studied.

Summarizing, the above attempts to cure the problem of stability have the general shortcoming,
that is, not complete account of all fluctuations leading to the existence of inter-component
as well as intra-component correlation functions, including all normal correlation functions
as well as the so-called anomalous averages. The latter often are neglected. Actually, in the
system with broken gauge symmetry, such as the system of bosons at very low temperatures,
anomalous averages play a very important role. At temperatures $T\leq T_c$ in the dilute Bose
gas with the critical temperature $T_c$, the absolute value of the anomalous average (density),
$|\sigma|$ can be of the same order as the normal density, $\rho_1$ or even larger
\cite{Yukalov_2005,Yuk_Yuk_2014,oursigma,Yukalov_2025}. Taking into account the anomalous
density is principally important for Bose-Einstein condensation in any system. This has been
confirmed also for the triplon condensation in quantum magnets. Particularly, it has been
shown that, neglecting $\sigma$ leads to an unphysical jump in magnetization curves
\cite{ourannals1}.

The  physical meaning of the normal density is straightforward: $\rho_1$ is the density
of uncondensed particles, with the normalization $\rho_0 +\rho_1=\rho$. As to the physical
meaning of $\sigma$, its absolute value, $|\sigma|$ gives the amplitude of pair processes,
when two particles are annihilated from the thermal cloud of non -- condensed particles.
In other words $|\sigma|^2$ describes the density of binary correlated particles, so that
the number of correlated pairs equals $|\sigma|^2/2$.

In the present work, we show that the problem of the complex sound velocity, when $c_d^2\leq0$,
can be solved by accurately taking into account all anomalous as well as mixed densities.
We employ optimized perturbation theory (OPT), whose idea was first proposed in
Ref. \cite{yukbull} and recently reviewed in Refs. \cite{Yukalov_2019,Yukalov_2021}.
This approach has been successfully applied for atomic gases \ci{ourtan1},
optical lattices \ci{ourlatiman,ourlatzabar}, and quantum magnets \cite{ourmce}.
Recently, we have developed OPT for a two-component Bose mixture with repulsive interactions,
with $(g_{aa}>0$, $g_{bb}>0$, and $g_{ab}>0)$ \cite{ourmix1}. In the present work, this
self -- consistent theory of a homogeneous binary Bose mixture will be applied to the case
of attractive inter-species interaction, where $(g_{aa}>0$, $g_{bb}>0$, but $g_{ab}<0)$.

The paper is structured as follows. In Sect. II we
derive general expressions for the energy, collective excitations spectrum and densities
in OPT. Then in Sect. III we apply this theory to symmetric Bose mixture with $g_{ab}<0$
to separate its stable and unstable regions. In Sect. IV we present the phase diagram of
the system on $(\alpha,\gamma)$ plane at $T=0$. The last section includes our conclusions
and discussions. In order not to overload the main part of the paper, clarifying important
explanations are shifted to Appendices.

\section{Optimized perturbation theory for a symmetric binary Bose mixture }

In the present section, we generalize the approach, developed in Ref. \cite{ourmix1} for
a binary Bose mixture, where all interactions are repulsive, to the binary mixture, where
inter-component interactions are attractive. Our approach takes into account all correlation
functions that exist in the generalization of the HFB method to binary mixtures, including
normal, anomalous and cross-component correlations. The divergences in the anomalous averages
are regularized in the standard way \cite{andersen} by employing the dimensional
regularization, which yields finite values and does not contain any cutoffs.

The Lagrangian density for two-species complex scalar fields $\psi$ and $\phi$, with contact
self-couplings $g_a$ and $g_b$ and inter-species coupling $g_{ab}$, is given as
\begin{eqnarray}
%\bea
%\begin{aligned}
	&& L=\psi^{\dagger} (i\partial_t+\frac{\bf{\nabla}^2}{2m_a}+\mu_a)\psi-\frac{g_a}{2}(\psi^{\dagger}\psi)^2+\phi^{\dagger} (i\partial_t + \\
	&& \frac{\bf{\nabla}^2}{2m_b}+\mu_b)\phi-\frac{g_b}{2}(\phi^\dagger\phi)^2-g_{ab}(\psi^\dagger\psi)(\phi^\dagger\phi) \; ,
	\label{Lagrange}
%\end{aligned}
%\eea
\end{eqnarray}
where the associated chemical potentials are represented by $\mu_{a,b}$, while $m_{a,b}$
represent the masses. In terms of the corresponding $s$-wave scattering lengths, the
coupling constants can be written as $g_{a,b}=4\pi a_{a,b}/m_{a,b}$, while the cross
coupling is $g_{ab}=2\pi a_{ab}/m_{ab}$, where $m_{ab}=m_am_b/(m_a+m_b)$ represents the
reduced mass. In quantum field theory, the grand canonical thermodynamic potential $\Omega$
of a statistical system in the equlibrum can be found by evaluating the partition function
$Z$ through the relation:
\begin{equation}
	\Omega=-T\ln{Z} \; ,
	\label{10.1}
\end{equation}
\begin{equation}
	Z= \dsint D\psi^\dagger D\psi D\phi^\dagger D\phi e^{-S[\psi^\dagger,\psi,\phi^\dagger,\phi]} \; ,
	\label{10.2}
\end{equation}
where the equivalent finite temperature Euclidean $(\tau=it)$ space time action is given by
%\bea
%\begin{aligned}
\begin{eqnarray}
	S=&\int_0^{\beta} d\tau \int d\vec{r} \left\lbrace \psi^{\dagger} \hat{K}_{a}\psi +\phi^{\dagger}\hat{K}_{b}\phi +\frac{g_a}{2}(\psi^{\dagger} \psi)^2\right. \\
	&+\left.\frac{g_b}{2}(\phi^{\dagger} \phi)^2+g_{ab}(\psi^{\dagger} \psi)(\phi^{\dagger} \phi)\right\rbrace, \\
	\hat{K}_{a,b}&=\frac{\partial}{\partial\tau}-\hat{O}_{a,b}; \ \ \ \ \ \ \  \ \hat{O}_{a,b}=\frac{\vec{\nabla}^2}{2m_{a,b}}+\mu_{a,b} \; .
	\label{10.3}
%\end{aligned}
%\eea
\end{eqnarray}
In Eq.(\ref{10.3}) the fields $\psi({\bf{r}},\tau)$ and $\phi({\bf{r}},\tau)$ are periodic
in $\tau$ with the period $\beta=1/T$. Clearly, due to the last term in (\ref{10.3}) the path
integral in (\ref{10.2}) cannot be evaluated exactly.

In his pioneering work, Petrov used the Bogolubov approximation, neglecting the fluctuating fields
$\tilde{\psi}$ and $\tilde{\phi}$ in the shift
\begin{equation}
	\ba
	\psi(\mathbf{r},\tau)=\sqrt{\rho_{0a}}+\tilde{\psi}(\mathbf{r},\tau),
	\quad \quad
	\phi(\mathbf{r},\tau)=\sqrt{\rho_{0b}}+\tilde{\phi}(\mathbf{r},\tau) \; ,
	\ea
	\label{10.4}
\end{equation}
Afterwards, the LHY term has been included "by hand". Actually, this approximation can
be consequently realized in the framework of a field theoretical approach referred as
bilinear approximation, which takes into account only quadratic terms in fluctuating fields.
As it is expected, this approximation, which also neglects anomalous densities, is
not able to improve the situation related to the complex sound velocity (see Appendix A).
Below we go beyond this approximation and revise the problem again.

The derivation of higher approximations above the bilinear approximation can be done by
employing the variant of perturbation theory called optimized perturbation theory (OPT)
\cite{Yukalov_2019,Yukalov_2021}. Recently we have developed OPT \cite{ourmix1}
for a two-component Bose system and studied its properties in the repulsive
($g_{a}>0, g_{b}>0, g_{ab}>0$) regime. Below, referring the reader to our previous work
\cite{ourmix1} for details, we consider the main equations for the symmetric case.

First, it should be emphasized that, in the present approach, for the system with Bose-Einstein
condensate, two kinds of chemical potentials are introduced instead of a unique chemical 
potential $\mu$, so that $\mu_{0a}N_{0a}+\mu_{1a}N_{1a}=\mu_{a}N_{a}$. The reason is the 
following. As is known, the theory of BEC has a long-standing problem, referred to as the 
Hohenberg-Martin dilemma \cite{HohenbergM}, who showed that in the standard approach 
with a single chemical potential either there appears a gap in the spectrum or the system 
is not thermodynamically stable. This problem is solved \cite{Yukalov_2005} by introducing 
two chemical potentials.  

According to the general theory of systems with broken global gauge symmetry, the number 
of Lagrange multipliers has to be equal to the dimensionality of the order parameter.
A system with Bose-Einstein condensate is necessarily accompanied by the global gauge symmetry 
breaking classified by O(2) symmetry, hence, characterized by two parameters. These symmetry 
specifying parameters are the amplitude of the condensate wave function and its phase,
which requires the introduction of two chemical potentials \cite{Yukalov_2005}. 

From the point of view of physics, there are two conditions imposed on the system, as soon 
as gauge symmetry is broken. For each of the components, the particle spectrum should satisfy 
the Goldstone theorem, which imposes a constraint on the chemical potential. The other 
constraint is the necessity for the thermodynamic potential to correspond to the minimum 
with respect to the condensate fraction. It has been shown that, when anomalous density is 
accurately taken into account, these two conditions cannot be satisfied simultaneously, but 
the introduction of two chemical potentials makes the theory self-consistent 
\cite{yukalovannals,ourAniz1}. Naturally, in the normal phase, when $\rho_0=0$, $\sigma=0$, 
both chemical potentials coincide: $\mu=\mu_0=\mu_1$. Mathematical justifications making the 
approach self-consistent have been detailed in Refs. \cite{Yukalov_2005,yukalovannals,Yukalov_2025}. 
To remind the reader the basic points of the approach, we add Appendix B. It is also important 
to emphasize that the developed approach has been applied to single-component systems with 
Bose-Einstein condensate, and, comparing the results with Monte Carlo numerical calculations 
\ci{Rossi}, the high accuracy of the approach has been demonstrated. The related discussion is 
given in Appendix C.

Thus, starting from the action \re{10.3}, OPT leads to the following inverse propagator
in the momentum space:

\begin{equation}
	G^{-1}(\omega_n,\bfk)=
	\begin{pmatrix}
			\veps(k)+X_1 & \omega_n & X_5 &0 \cr
		-\omega_n    &\veps(k)+X_2 & 0 & 0 \cr
		X_5 & 0 &\veps(k)+X_1 &\omega_n \cr
		0 & 0 &-\omega_n &\veps(k)+X_2 \cr 
\end{pmatrix}
\label{15.1}
\end{equation}

Here, due to the Hugenholtz-Pines relations, we have $X_2=X_6=0$, while other variational
parameters $X_1\equiv 2\Delta_1$ and $X_5\equiv 2\Delta_{12}$ are fixed by the minimum of
the thermodynamic potential $\Omega$ , given by the Eq. (43) of Ref. \cite{ourmix1}.
Equating the determinant of $G^{-1}$ to zero, one obtains two branches of energy dispersion
\begin{equation}
	\begin{array}{cc}
		&  \omega_{\pm}^2=\omega_{d,s}^2=\varepsilon_k
		(\varepsilon_k+2mc_{\pm}^2) ,
			\end{array}
	\label{16.4}
\end{equation}
where the sound  velocities are related to the self energies as
\begin{equation}
	c_d^2\equiv c_+^2=\dsfrac{\Delta_1+\Delta_{12}}{m} ,\quad \quad
	 c_s^2\equiv c_-^2=\dsfrac{\Delta_1-\Delta_{12}}{m} \; .
	\label{16.5}
\end{equation}
It is pleasant to note that, in the present theory both branches of the collective excitations
in BEC phase are gapless. The minimization of the corresponding thermodynamic potential leads
to the following equations with respect to the variational parameters:
\footnote{Neglecting in \re{16.6} $\sigma_a$ and $\rho_{ab}$
one comes back to the Eq. \re{a13shtrix3} of bilinear approximation.}
\begin{equation}
	\begin{array}{cc}
		&  \Delta_1=g[\rho_{0a}+\sigma_a] \; ,\\
		& \Delta_{12}=\dsfrac{g_{ab}}{2}(2\rho_{0a}+  \rho_{ab}) \; ,
	\end{array}
	\label{16.6}
\end{equation}
where the number density of condensed particles, say, of the type $a$ satisfies
the normalization condition $\rho_{0a}=\rho_{a}-\rho_{1a}$. Other densities are defined as
\be
\ba
\rho_{1a}=\dsfrac{N_{1a}}{V}=\int d\bfr
\langle\tilde{\psi}^\dagger(\bfr)\tilde{\psi}(\bfr)\rangle =\frac{1}{2V}
\int d\bfr[G_{11}(\bfr,\bfr)
+G_{22}(\bfr,\bfr)], \\
\\
\sigma_a=\dsfrac{1}{2V} \int d\bfr[\langle \Tilde{\psi}^\dagger(\bfr)\Tilde{\psi}^\dagger(\bfr)\rangle
+ \langle \Tilde{\psi}(\bfr)\Tilde{\psi}(\bfr)\rangle]=
\int d\bfr[G_{11}(\bfr,\bfr)
-G_{22}(\bfr,\bfr)], \\
\\
\rho_{ab}  =\dsfrac{1}{V} \int d\bfr[\langle \Tilde{\psi}^\dagger(\bfr)\Tilde{\phi}(\bfr)\rangle + 
\langle \Tilde{\phi}^\dagger(\bfr)\Tilde{\psi}(\bfr)\rangle]
=\dsfrac{1}{V}\int d\bfr[G_{13}(\bfr,\bfr)+G_{24}(\bfr,\bfr)].
\label{n1a1b}
\ea
\ee
At zero temperature, for a uniform system, one can evaluate these integrals analytically in
the momentum space:
\be
G_{ij}(\bfr,\tau;\bfr',\tau')=\dsfrac{1}{V\beta}\dssum_{n,k}
e^{i\bfk(\bfr-\bfr')}e^{i\omega_n(\tau-\tau')}G_{ij}(\omega_n, \mathbf{k} )\vert_{\tau=\tau',\bfr=\bfr'}\; ,
\lab{momg}
\ee
to obtain
\nopagebreak
%\begin{widetext}
	\be
	\ba
	\rho_{1a}= \dsfrac{1}{2V}\sum_k\{\dsfrac{\varepsilon_k+\Delta_1
		+\Delta_{12}}{2\omega_d}+\dsfrac{\varepsilon_k+\Delta_1-\Delta_{12}}{2\omega_s}-1\}=
	\dsfrac{m^3(c_d^3+c_s^3)}{6\pi^2}\equiv\dsfrac{\rho_1}{2} \; ,
	\label{17.1}
	\ea
	\ee
	\begin{equation}
		\ba
		\sigma_a= -\dsfrac{1}{2V}\sum_k\{\dsfrac{\Delta_1+\Delta_{12}}{2\omega_d}-\dsfrac{\Delta_1-\Delta_{12}}{2\omega_s}-\dsfrac{\Delta_1}{\varepsilon_k}\}=
		\dsfrac{m^3(c_d^3+c_s^3)}{2\pi^2}\equiv\dsfrac{\sigma}{2} ,
		\label{17.1s}
		\ea
	\end{equation}
	%\nopagebreak
	%\end{widetext}
	\begin{equation}
		\ba
		\rho_{ab}= \dsfrac{1}{2V}\sum_k\{\dsfrac{\varepsilon_k+\Delta_1+\Delta_{12}}{\omega_d}-\dsfrac{\varepsilon_k+\Delta_1-\Delta_{12}}{\omega_s}\}=
		\dsfrac{m^3(c_d^3-c_s^3)}{3\pi^2} ,
		\label{17.2}
		\ea
	\end{equation}
%\end{widetext}
where we used Eqs. (\ref{16.5}). Now, using Eqs. (\ref{16.5}) - (\ref{17.2}),
one obtains the system of two nonlinear algebraic equations with respect to the sound velocities:
\begin{equation}
	c_d^3-\dsfrac{3\pi^2 c_d^2}{gm^2}+(2+\tilde{\alpha}^2)c_s^3-\dsfrac{3\pi^2\tilde{\alpha}^2\rho}{2m^3}=0 \; ,
	\label{17.3}
\end{equation}
\begin{equation}
	c_d^3-\frac{3\pi^2c_s^2}{gm^2}-\tilde{\alpha}^2c_s^3
	+\frac{3\pi^2\rho(2+\tilde{\alpha}^2)}{2m^3}=0 \; .
	\label{17.4}
\end{equation}
Here keeping in mind the negative sign of $g_{ab}$, we introduce the notation
\begin{equation}
	{\tilde\alpha}^2=-\delta g/g=-\alpha.
	\label{16.7}
\end{equation}
These equations can be rewritten in the dimensionless form as
\begin{equation}
    s_d^3-\dsfrac{3\pi}{4}s_d^2+(2+\tilde{\alpha}^2)s_s^3-\dsfrac{3\pi^2\gamma\tilde{\alpha}^2}{2}=0 \; ,
    \label{17.5}
\end{equation}
\begin{equation}
    s_d^3-\tilde{\alpha}^2s_s^3-\dsfrac{3\pi s_s^2}{4}+\dsfrac{3\pi^2\gamma(2+\tilde{\alpha}^2)}{2}=0 \; ,
    \label{17.6}
\end{equation}
where $s_{d,s}=c_{d,s}ma_s$ and $\gamma$ is the gas parameter $\gamma=\rho a_s^3$. Of course,
omitting here the contributions from $\sigma_a$ and $\rho_{ab}$ one would come back to
the Eqs. (\ref{a13.2}) of the bilinear approximation, outlined in the Appendix A. Hopefully,
the presence of additional terms
in (\ref{17.5}) and (\ref{17.6}), due to $\sigma$ and $\rho_{ab}$, will facilitate the solution
of the problem of a complex sound velocity, when $c_d^2<0$. From the last equations it
is seen that at $T=0$ the sound velocities for symmetric binary mixture depend only on the two
independent parameters, $\gamma$ and $\tilde{\alpha}^2=-\delta g /g=-(g_{ab}+g)/g\ge0$ \; .

Clearly, the solutions to these equations have to satisfy the following physical conditions: \\
1) Positivity of the sound velocities, $s_d\ge 0$, $s_s\ge0$ \; .\\
2) Normalization condition: $\rho_{1a}/\rho_a\leq1$ i.e.
\be
\rho_{1a}=\frac{s_d^3+s_s^3}{6\pi^2a_s^3}\leq \frac{\rho}{2} \; ,
\lab{normdense}
\ee
 (see Eq.(\ref{17.1})), and  hence
\begin{equation}
    \frac{s_d^3+s_s^3}{3\pi^2}\leq \gamma \; .
\end{equation}
In the next section we shall analyze the solutions to Eqs. (\ref{17.5}) and (\ref{17.6})
numerically and study the phase diagram of the system on $(\tilde{\alpha}^2,\gamma)$ plane.

\section{Region of stability in $(\tilde{\alpha}^2,\gamma)$ plane}

It is clear that, when the inter-species attraction is too strong, so that $|\delta g/g|>>1$
the system collapses. In terms of the energy dispersion this situation corresponds to the
dynamical instability, related to a complex sound velocity. The stable equilibrium can be
reached under the appropriate balance between competitive attractive and repulsive forces.
This implies that the stability condition defines a region on the plane connecting the
parameter $\gamma$, responsible for the repulsion, and $\tilde{\alpha}^2=-\delta g/g$,
responsible for the attraction.

To find the boundary of stability, we use Eqs. (\ref{17.5}) and (\ref{17.6}), requiring
that $s_d\ge0$. Since the lower boundary is $s_d=0$, we  set  $s_d=0$ in both equations
and solve the system of equations
\begin{equation}
 (2+\tilde{\alpha}^2)s_s^3-\frac{3\pi^2\tilde{\alpha}^2\gamma}{2}=0 \; ,
 \label{19.1}
\end{equation}
\begin{equation}
 \tilde{\alpha}^2s_s^3+\frac{3\pi s_s^3}{4}-\dsfrac{3\pi^2\gamma(2+\tilde{\alpha}^2)}{2}=0 \; ,
 \label{19.2}
\end{equation}
with respect to $\gamma$ and $s_s$ for a fixed $\tilde{\alpha}^2$. Then we obtain
\begin{equation}
 \gamma_{crit}=\dsfrac{9\tilde{\alpha}^4\pi(2+\tilde{\alpha}^2)}{2048(1+\tilde{\alpha}^2)^3}\ge0 \; ,
 \label{19.3}
\end{equation}
\begin{equation}
s_s=\dsfrac{3\tilde{\alpha}^2\pi}{16(1+\tilde{\alpha}^2)}\ge0 \; .
\label{19.4}
\end{equation}
Therefore, for a given $\delta g/g=-\tilde{\alpha}^2$, the symmetric binary Bose mixture
may be in a stable equilibrium state if $\gamma=\rho a_s^3\ge \gamma_{crit}$.
This is illustrated in Fig. 1a, where on the ordinate axis we present $-\log_{10}(\gamma)$,
since $\gamma$ may vary in a large scale. Thus, we show that, when $\gamma \ge \gamma_{crit}$,
the density sound velocity $c_d$ is positively defined.

On the other hand, the question arises whether there exists a relation between $\gamma$ and
$\delta g/g$ establishing the positiveness of the pseudo-spin sound velocity, $c_s=s_s/(a_s m)$.
To this end, let in Eqs. (\ref{17.5}) and (\ref{17.6}), $s_s=0$, so that we have
\begin{equation}
    \begin{array}{cc}
         &  s_d^3-\dsfrac{3\pi}{4}s_d^2-\dsfrac{3\pi^2\tilde{\alpha}^2\gamma}{2}=0 \; ,\\
         &  s_d^3+3\pi^2\gamma+\dsfrac{3\pi^2\tilde{\alpha}^2\gamma}{2}=0 \; .
    \end{array}
    \label{20shtrix1}
\end{equation}
Solving these equations with respect to $\gamma$ and $s_d$ we find
 \begin{equation}
     \begin{array}{cc}
          &  s_d=\dsfrac{3\pi(2+\tilde{\alpha}^2)}{8(1+\tilde{\alpha}^2)} \; ,\\
          &  \gamma_{crit}(s_s=0)=-\dsfrac{9\pi(2+\tilde{\alpha}^2)^2}{256(1+\tilde{\alpha}^2)^3}
          <0 \; .
     \end{array}
     \label{20shtrix2}
 \end{equation}
Thus, we see that, $s_s$ becomes negative only for negative values of $\gamma=\rho a_s^3$. In
other words when intraspecies interaction is repulsive, the spin sound velocity $s_s$ is
always positive, regardless of the sign of inter-species interaction. This pleasant feature
is in contrast to the undesirable property of $s_s$ encountered in Ref. \cite{astra}. In fact,
it can be easily shown (see equation  below) that, their pseudo-spin sound velocity
becomes even complex $(s_s^2<0)$ for
\begin{equation}
          \gamma>\dsfrac{9\pi(1+{\tilde{\alpha}}^2)^2}
          {128{\tilde{\alpha}}^4(2+{\tilde{\alpha}}^2)^3} \; .
          \label{20shtrix3}
\end{equation}

To consider the role of the anomalous density, we plot in Fig. 1b the phase diagram, when
$\sigma_a$ in Eqs. \re{16.6} is neglected. It is seen that, in this
case the area of the stability region becomes smaller, to completely vanish when both
$\sigma_a$ and $\rho_{ab}$ are neglected as in some models.

%%%%%%%%%%%%%%%%%%%%  FIGURE 1 
\begin{figure}[H]
	\begin{minipage}[H]{0.49\linewidth}
		\center{\includegraphics[width=1.1\linewidth]{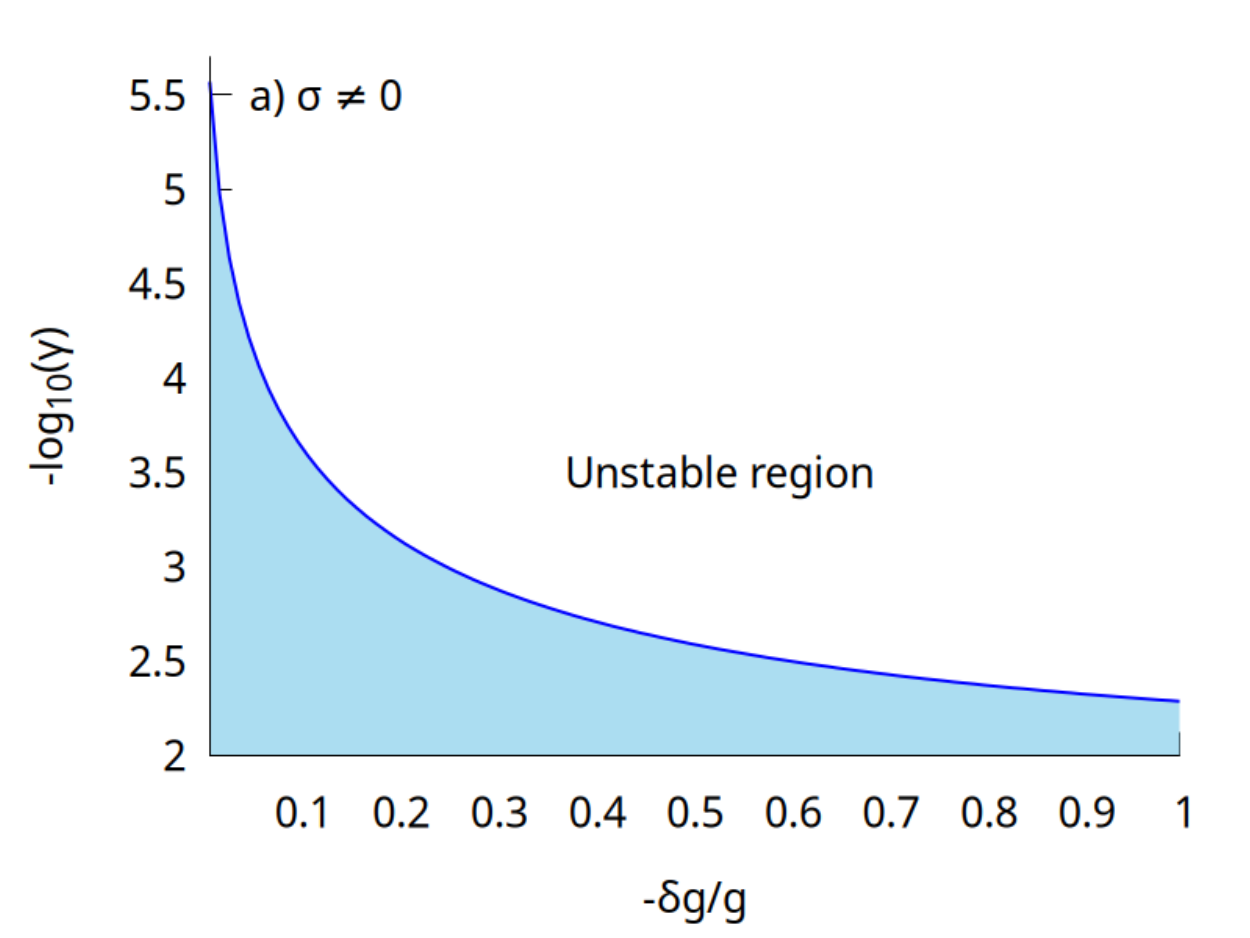} }
	\end{minipage}
	\hfill
	\begin{minipage}[H]{0.49\linewidth}
		\center{\includegraphics[width=1.1\linewidth]{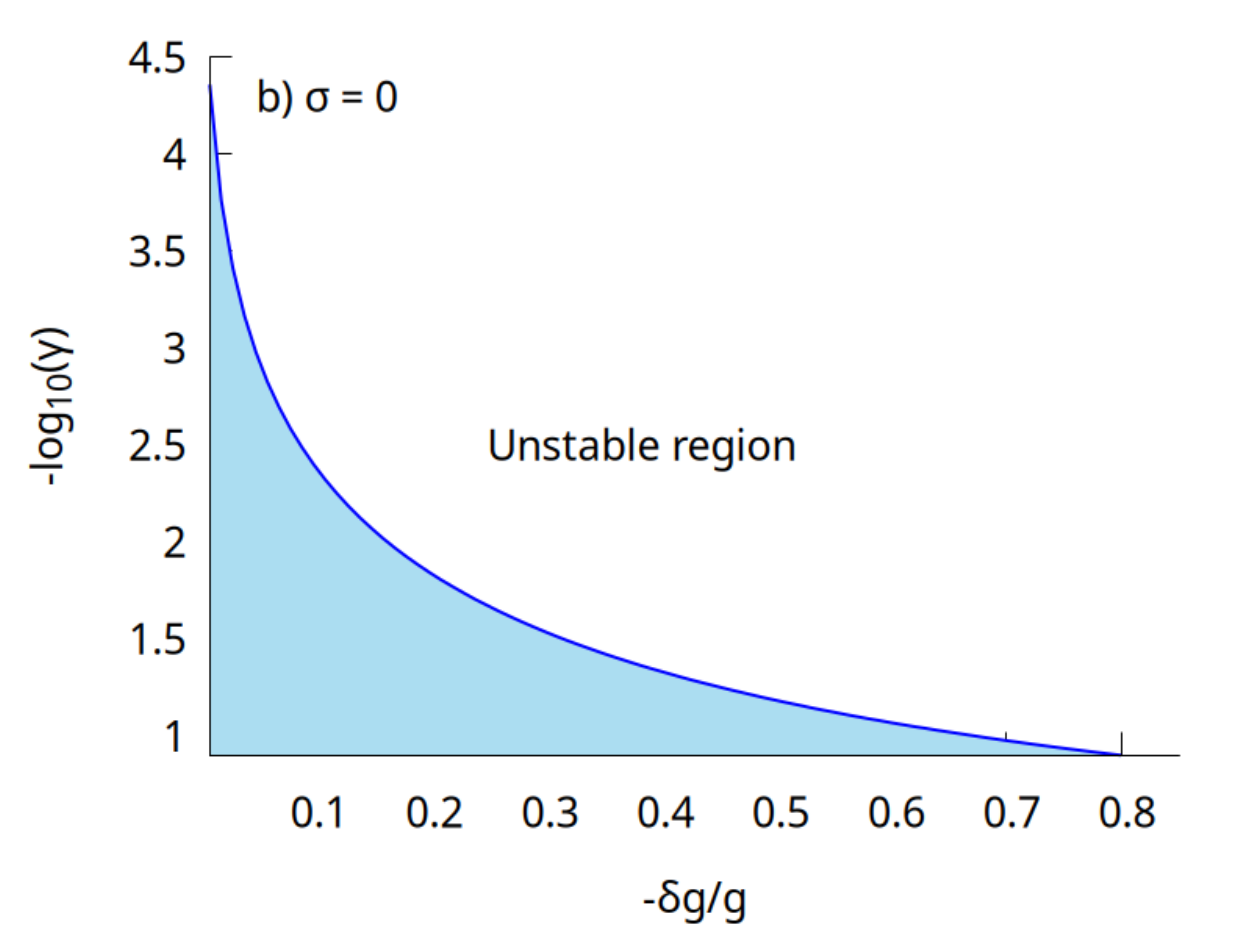} }
	\end{minipage}
	\caption{ (a): The phase diagram of symmetric binary Bose mixture on $(- \delta g /g,\gamma)$ plane; 
		(b): The same as in (a) but the anomalous density $\sigma$ is neglected }
	\label{Fig1}
\end{figure}

For further illustration  the importance of the anomalous and mixed densities, we present in Figs. 2
the fraction of condensed atoms (solid line), the anomalous (dashed line) and the mixed
densities, denoted by $n_0=\rho_{0a}/(\rho/2)$, $m=\sigma_a/(\rho/2)$ and
$n_{12}=\rho_{ab}/(\rho/2)$, respectively. It is seen that, $\sigma$ is positive, while
$\rho_{ab}$ is negative in the range of $\gamma\sim10^{-4}\div 10^{-1}$ and
$-\delta g/g \sim 0.1\div 0.7$. Moreover the absolute value of $\sigma$ is, in general,
of the same order as the order of the condensed fraction $\rho_0$.

The negative sign of $\rho_{ab}$, in accordance with Eq. (\ref{17.2}), is caused by the
fact that, $c_d<c_s$, as it is predicted in the Petrov's paper \cite{petrov}.
To illustrate this fact quantitatively, we present in  Fig. 3 the dimensionless sound
velocities $s_d$ and $s_s$ for three values of $-\delta g/g$. It is seen that $s_d/s_s$
decreases with the increase of attraction amplitude $|\delta g/g|$.

%%%%%%%%%%%%% FIGURE 2  %%%%%%%%%%%%%%%%%%%%%%%%%%
\begin{figure}[H]
	\begin{minipage}[H]{0.49\linewidth}
		\center{\includegraphics[width=1.1\linewidth]{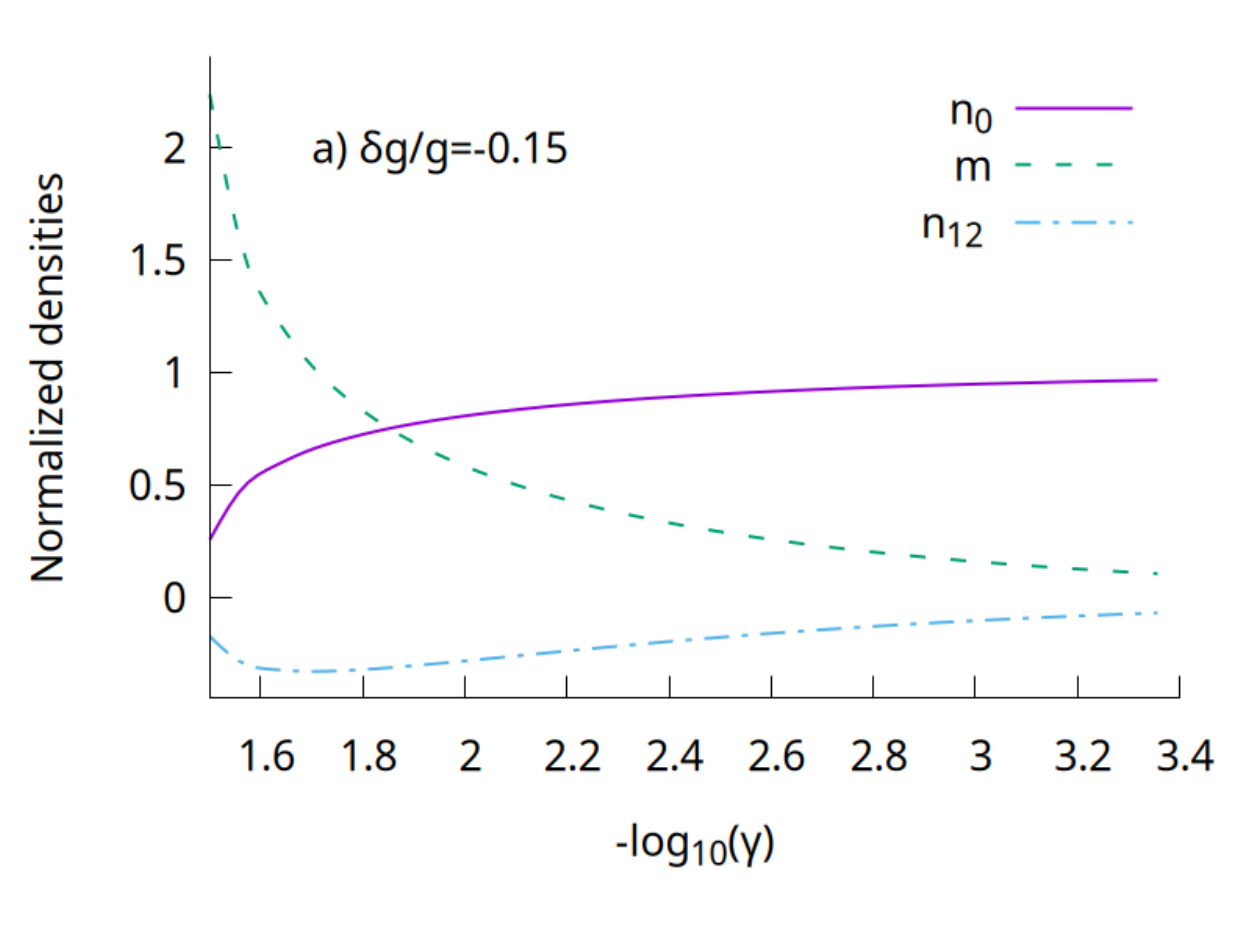} }
	\end{minipage}
	\hfill
	\begin{minipage}[H]{0.49\linewidth}
		\center{\includegraphics[width=1.1\linewidth]{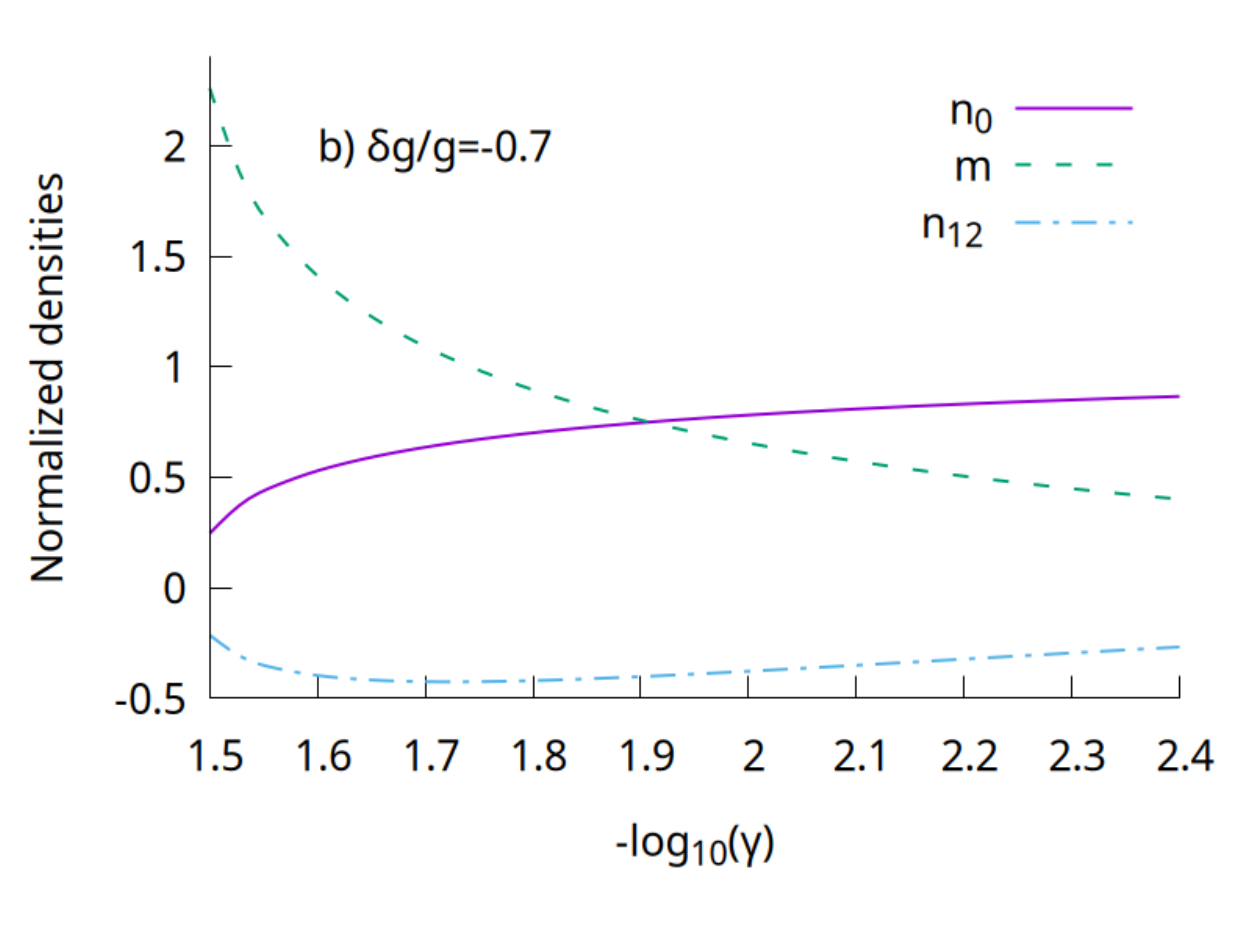} }
	\end{minipage}
	
	\caption{ The fraction of condensed atoms (solid line), the anomalous (dashed 
		line) and the mixed (dotted line) densities, for: (a) $\delta g/g=-0.15 $ and 
		(b) $\delta g/g=-0.7 $. 
	}
	\label{Fig2}
\end{figure}

%%%%%%%%%%%%%  FIGURE 3
\begin{figure}[H]
	\begin{minipage}[H]{0.32\textwidth}
		\center{\includegraphics[width=\columnwidth]{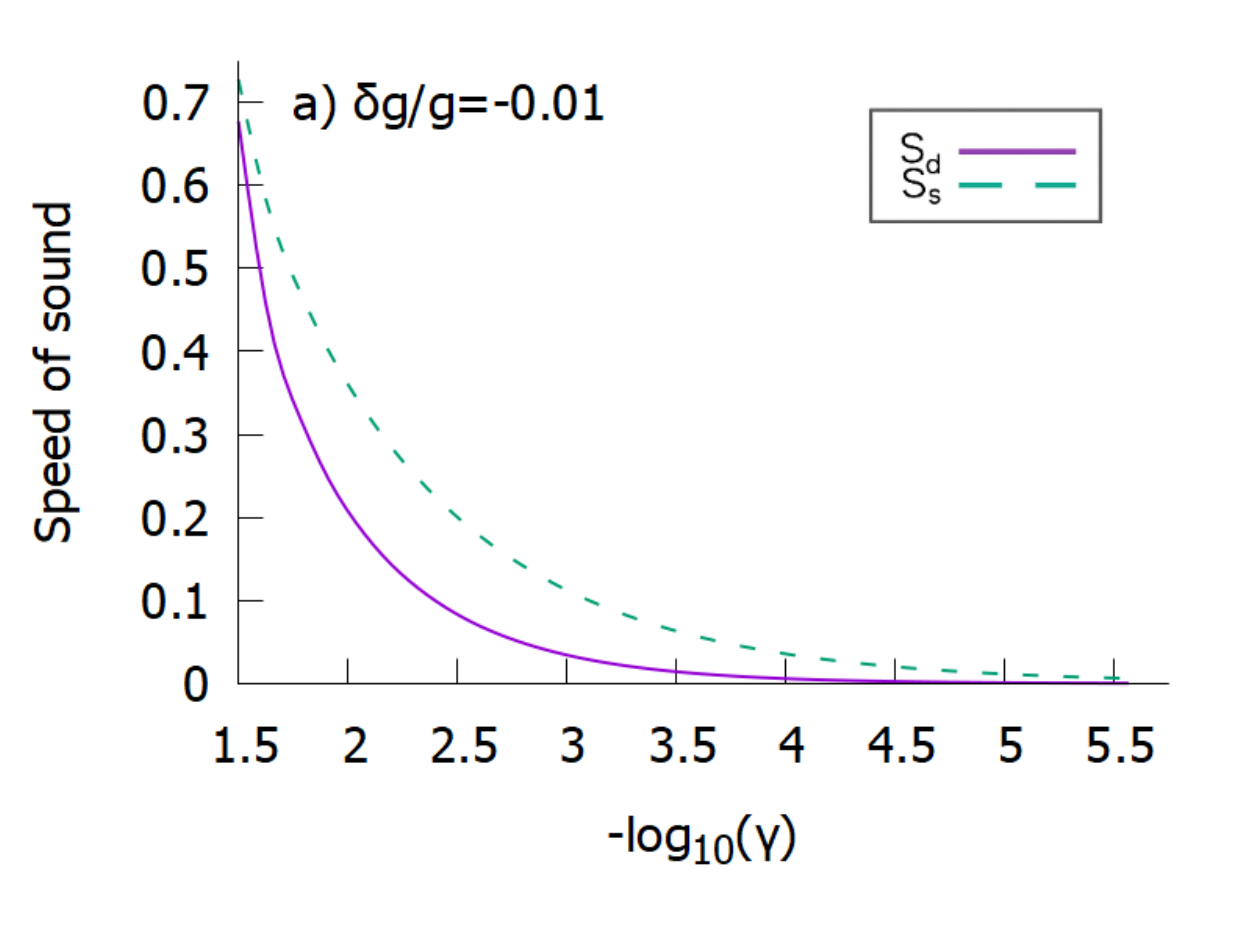} }
	\end{minipage}
	\hfill
	\begin{minipage}[H]{0.32\textwidth}
		\center{\includegraphics[width=\columnwidth]{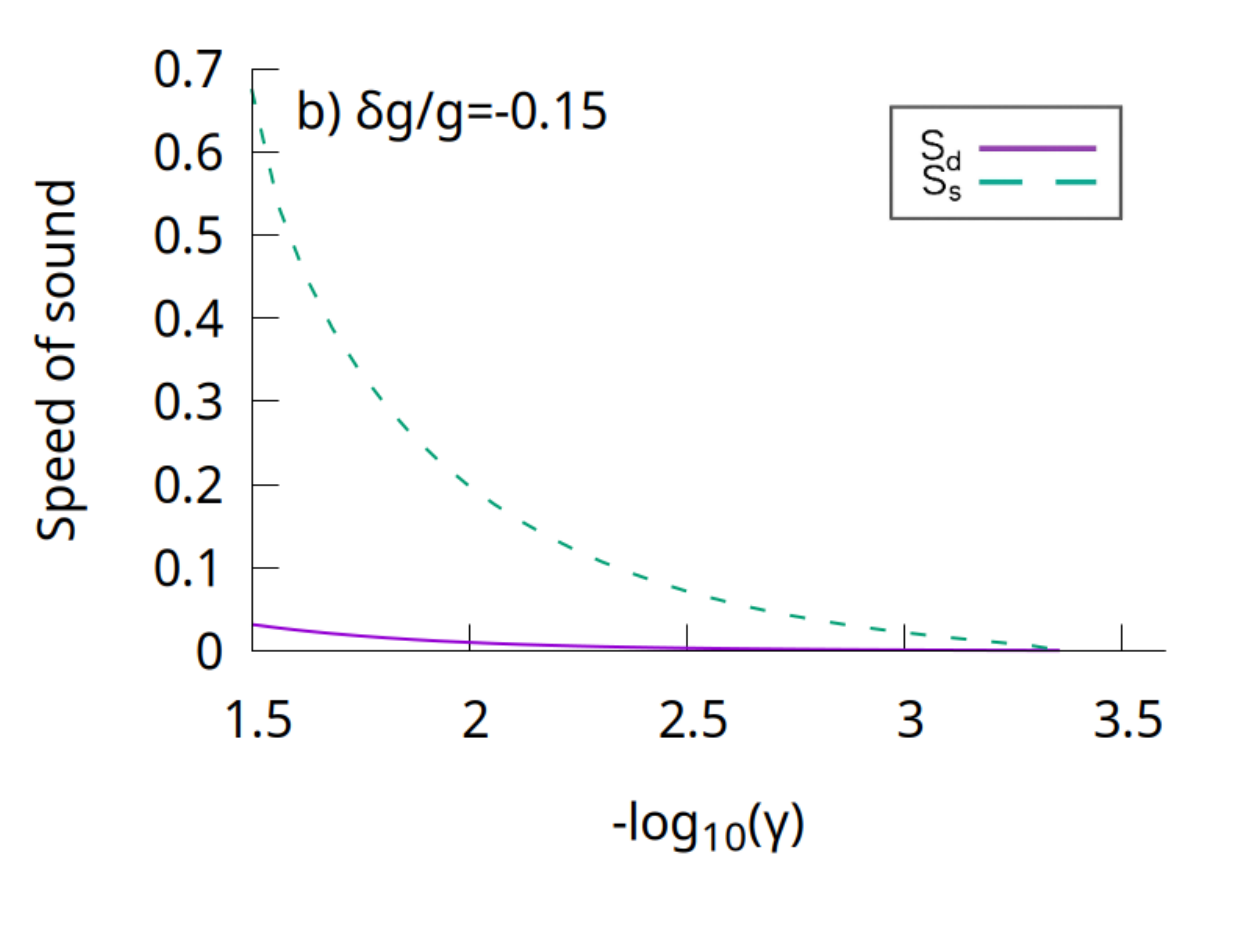} }
	\end{minipage}
	\begin{minipage}[H]{0.32\textwidth}
		\center{\includegraphics[width=\columnwidth]{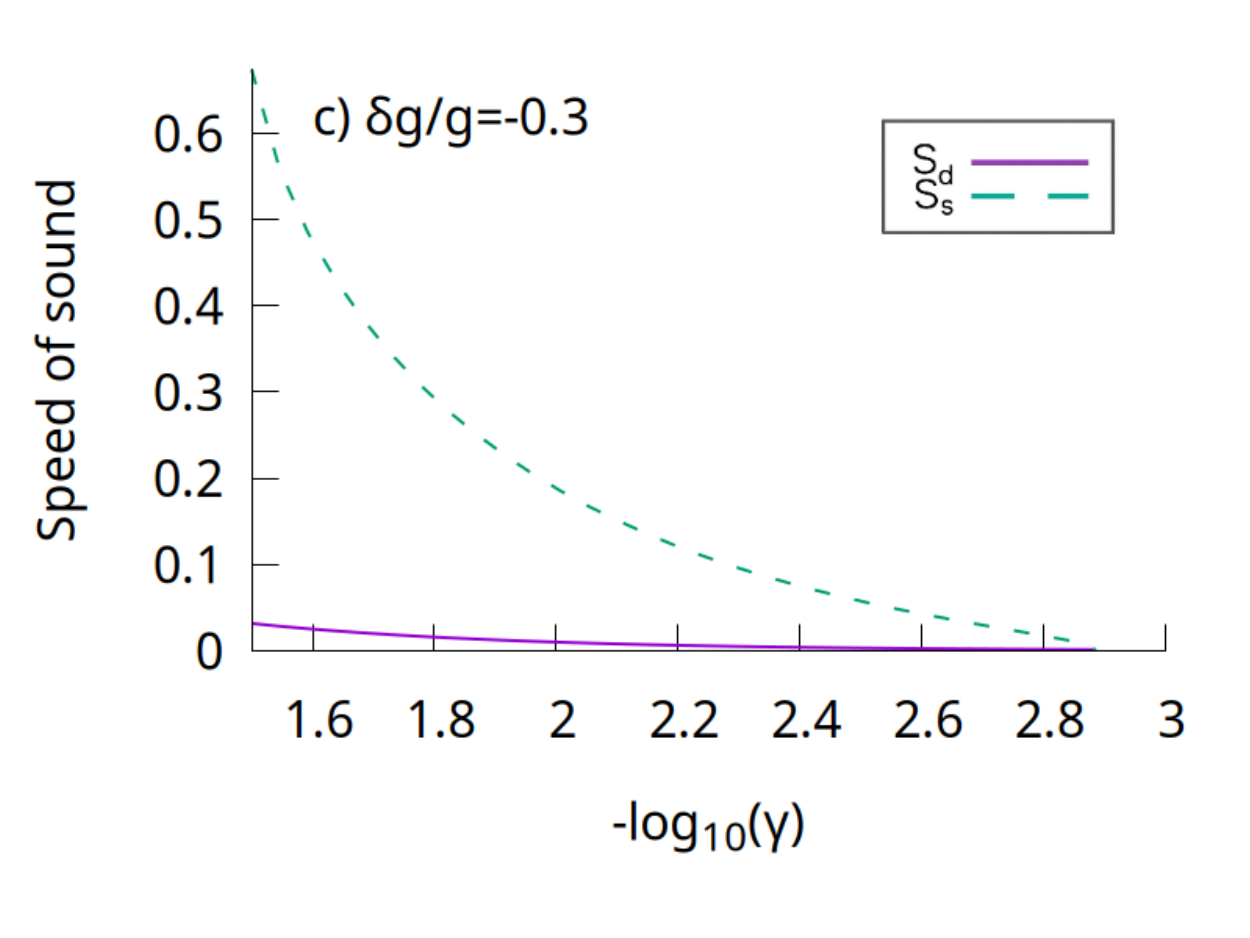} }
	\end{minipage}
	\hfill

	\caption{The dimensionless sound velocities $s_{p,m}=c_{p,m}ma_s$ vs 
		$-\log_{10}(\gamma)$ for: (a) $\delta g/g =-0.01$; (b) $\delta g/g=-0.15$, and (c) 
		$\delta g/g=-0.3$. }
	\label{Fig3}
\end{figure}

%%%%%%%%%%%%%  FIGURE 4
\begin{figure}[H]
	\begin{minipage}[H]{0.32\textwidth}
		\center{\includegraphics[width=\linewidth]{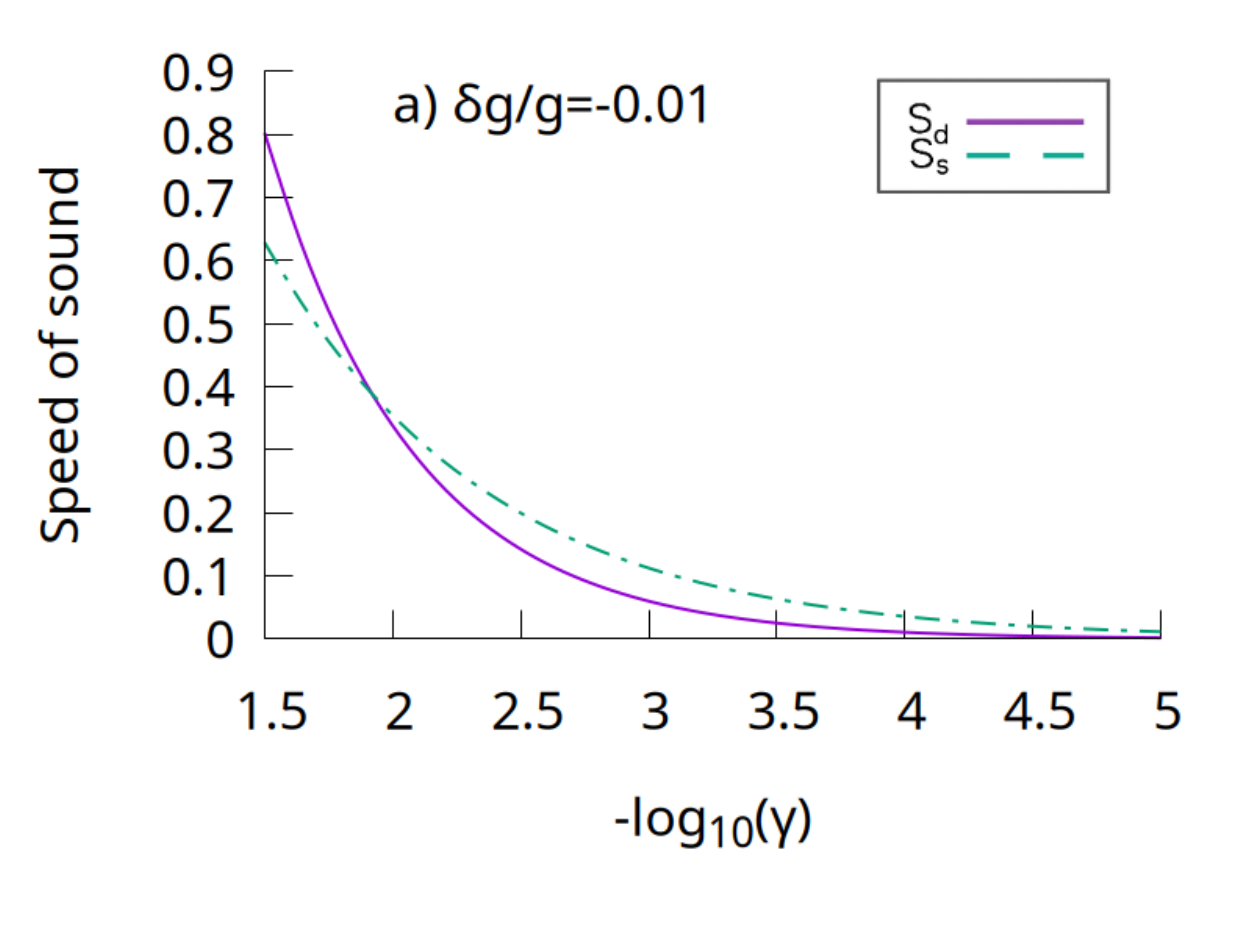} }
	\end{minipage}
	\hfill
	\begin{minipage}[H]{0.32\textwidth}
		\center{\includegraphics[width=\linewidth]{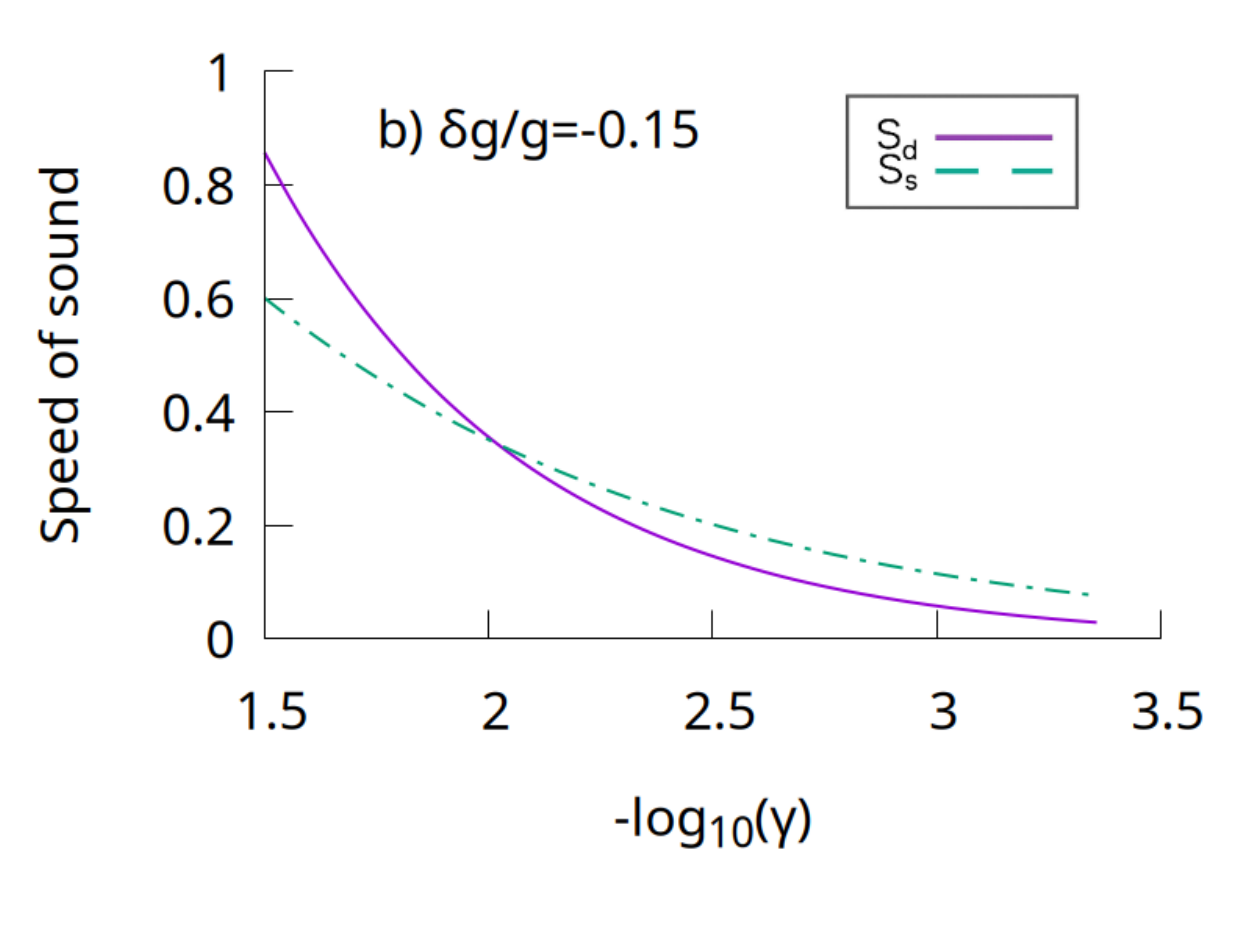} }
	\end{minipage}
	\begin{minipage}[H]{0.32\textwidth}
		\center{\includegraphics[width=\linewidth]{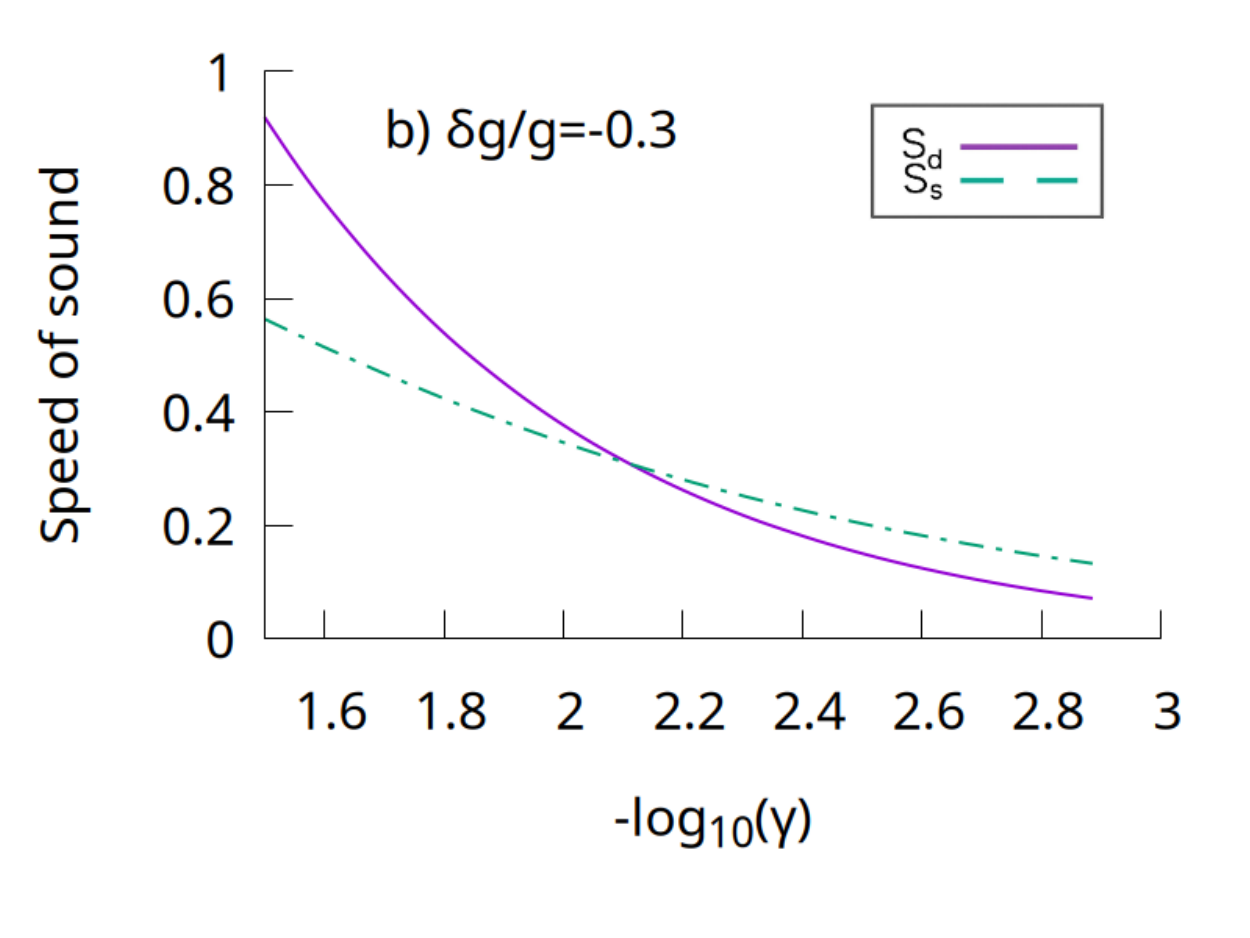} }
	\end{minipage}
	\hfill
	
	\caption{ The same as in Fig. 3  but in the model by Ota and Astrackharchik \ci{astra} }.
	\label{Fig4 }
\end{figure}

In Figs. 4(a,b,c), we present these sound velocities vs $-\log_{10}(\gamma)$ for
$\tilde{\alpha}^2=0.01$, $\tilde{\alpha}^2=0.15$, and $\tilde{\alpha}^2=0.3$. It is seen
that in this phenomenological model the condition $s_d\leq s_s$ is not always valid.
Unfortunately, experimental measurements of the sound velocities for unpolarized binary
Bose mixtures with attractive inter-species regime are not still available.

\section{Droplets vs dimerized gases}

 In the previous section, we have shown that, at $T=0$, the system can survive in a stable
equilibrium state when the condition (\ref{19.3}) relating $\gamma$ and $\delta g/g$ is
satisfied. This stable state can correspond to a liquid droplet or a gas of dimers, depending
on the sign of the total energy $\cale_{tot}$ of the system. In present theory $\cale_{tot}$
is given by Eq. (72) of Ref.\ci{ourmix1}:
 \begin{equation}
     \cale_{tot} =\cale_{GS}+\cale_{LHY}+\cale_{FLUC} \; .
 \end{equation}
 For the symmetric case, introducing the dimensionless energy per particle
 \begin{equation}
     \bar\cale_{tot}=\dsfrac{2\cale_{tot}ma_s^2}{N} \; ,
     \label{22.1}
 \end{equation}
 one obtains
 \begin{equation}
     \bar\cale_{tot}=\bar\cale_{GS}+\bar\cale_{LHY}+\bar\cale_{FLUC} \; ,
     \label{22.2}
 \end{equation}
 \begin{equation}
     \bar\cale_{GS}=-2\gamma \pi \tilde{\alpha}^2n_{0a}^2\leq0 \; ,
     \label{22.3}
 \end{equation}
 \begin{equation}
     \bar\cale_{LHY}=\dsfrac{16(s_d^5+s_s^5)}{15\gamma \pi^2}\ge0 \; ,
     \label{22.4}
 \end{equation}
 \begin{equation}
     \bar\cale_{FLUC}=\pi \gamma [(\tilde{\alpha}^2+1)(2n_{1a}^2+n_{ab}^2)-2m_a^2-2n_{1a}(m_a+\tilde{\alpha}^2)] \; ,
     \label{23.1}
 \end{equation}
where $n_{0a}=\rho_{0a}/(\rho/2)$, $n_{1a}=\rho_{1a}/(\rho/2)$, $m_a=\sigma_a/(\rho/2)$, and
$n_{ab}=\rho_{ab}/(\rho/2)$ are defined in Eqs. (\ref{17.1}). The first term in (\ref{22.2})
equals $\bar\cale_{GS}$ in the Bogoliubov approximation, the second term, $\bar\cale_{LHY}$
exactly coincides with the familiar expression given in the literature \cite{petrov,astra}.
While the third term, being quadratic in the densities, includes fluctuations as the
second order corrections.

%%%%%%%%%%%%%%%  FIGURE 5
\begin{figure}[H]
	\begin{minipage}[H]{0.32\textwidth}
		\center{\includegraphics[width=\linewidth]{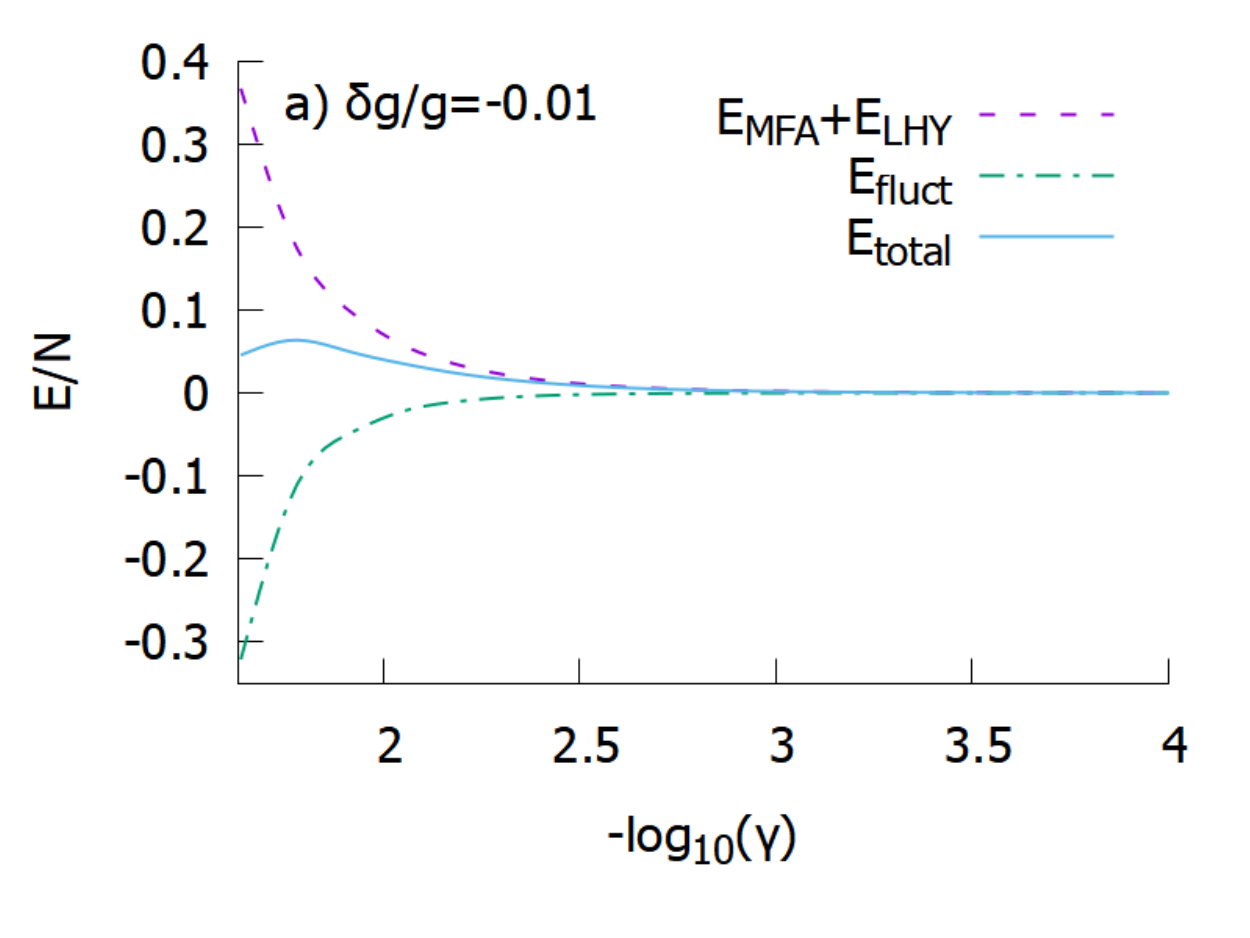} }
	\end{minipage}
	\hfill
	\begin{minipage}[H]{0.32\textwidth}
		\center{\includegraphics[width=\linewidth]{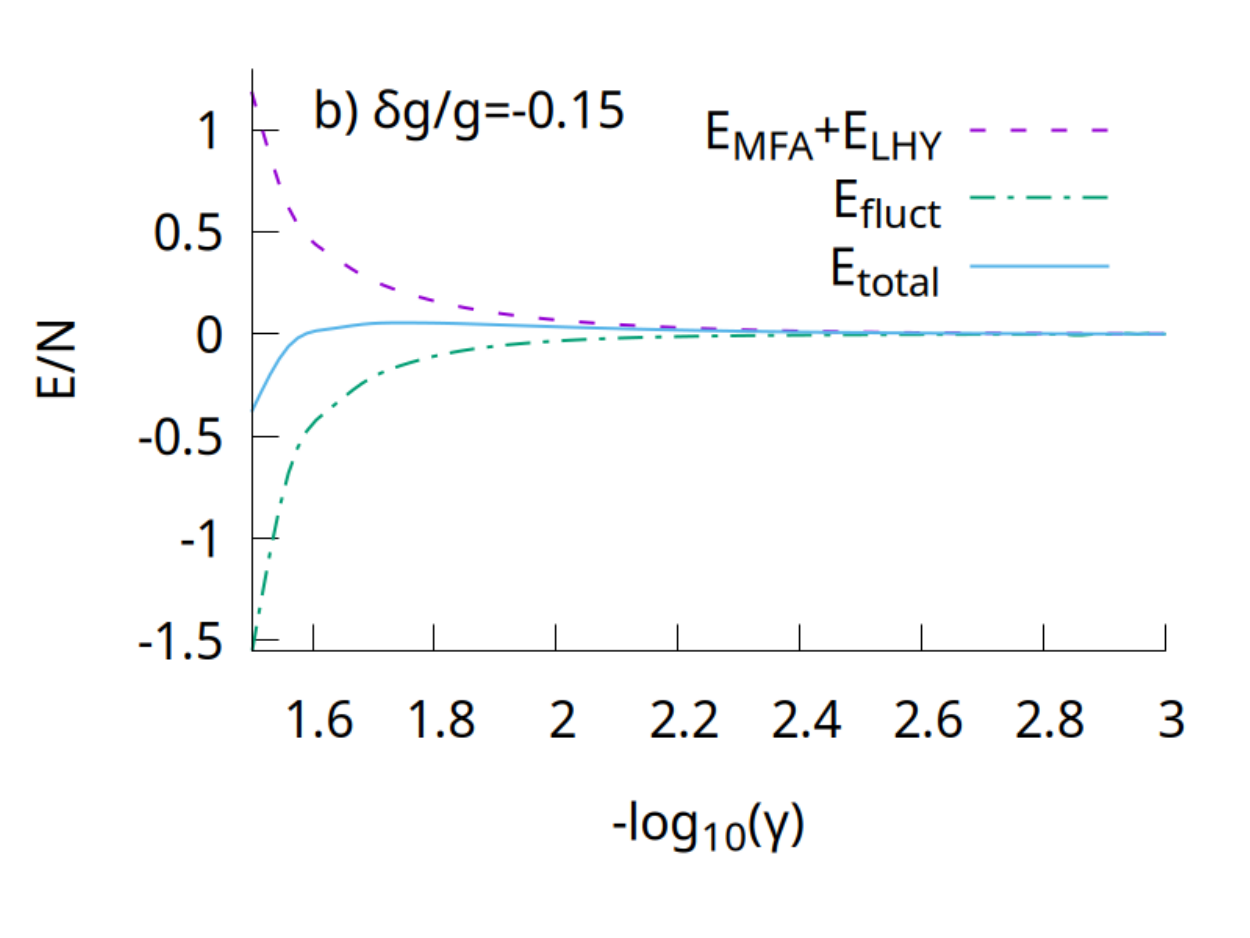} }
	\end{minipage}
	\begin{minipage}[H]{0.32\textwidth}
		\center{\includegraphics[width=\linewidth]{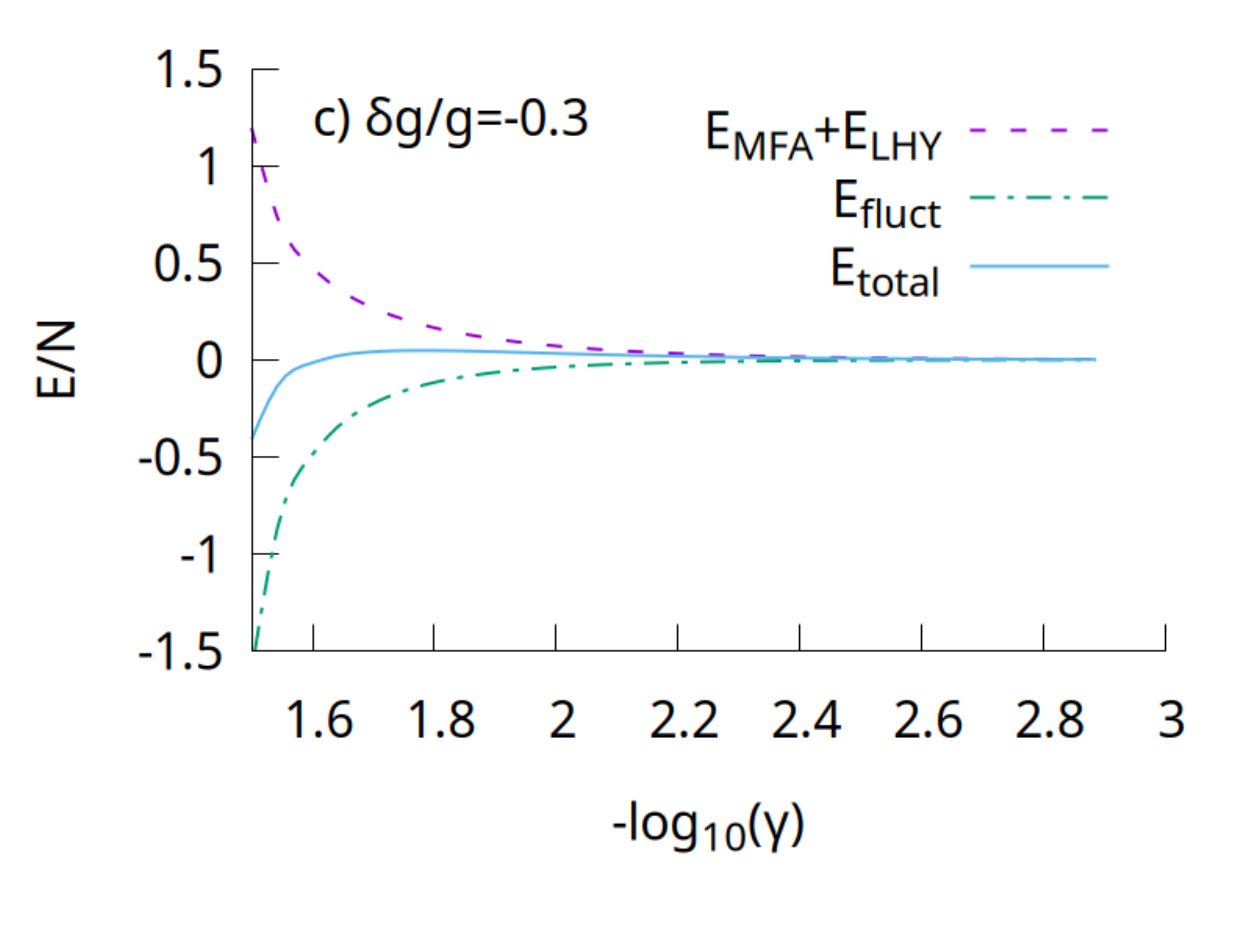} }
	\end{minipage}
	\hfill
	\caption{The dimensionless total energy of the system per particle for: 
		(a) $\delta g/g =-0.01$, (b) $\delta g/g=-0.15$, and (c) $\delta g/g=-0.3$}
	\label{Fig5}
\end{figure}

In Fig. 5 we plot the dimensionless energy for three values of $\tilde{\alpha}^2$, for
$\delta g/g =-0.01$, $\delta g/g=-0.15$, and $\delta g/g=-0.3$ vs $-\log_{10}(\gamma)$
in the region of stability. It is seen that the contribution from the second order
corrections, $\bar\cale_{fluc}$ (dotted lines) is always negative, which is favorable
for the droplet formation. However, due to the repulsive term $\bar\cale_{LHY}$ the total
energy remains positive for a weak attraction $\delta g/g=-0.01$  (see Fig. 5a). To make
the whole energy (solid lines) negative, one has to increase the intensity of the
inter-species attraction, as it is illustrated in Fig. 5b and Fig. 5c.

\subsection{Phase diagram of a liquid - gas state}

Without going into details, let us assume that $\bar\cale_{tot}<0$ and $\bar\cale_{tot}>0$
cases in the stability region, shown in Fig.1a, correspond to the droplet and gaseous phases,
respectively. Then the boundary of the phases lies on the points
$(\gamma=\gamma_0, \tilde{\alpha}^2=\tilde{\alpha}_0^2)$ where
\begin{equation}
    \label{24.1}
    \bar\cale_{Tot}(s_d,s_s,\gamma,\tilde{\alpha}^2)|_{\gamma=\gamma_0,\tilde{\alpha}^2=\tilde{\alpha}^2_0}=0 \; .
\end{equation}
Therefore, for each fixed value of $-\delta g/g=\tilde{\alpha}^2$, we have the system of
three nonlinear algebraic equations: (\ref{17.5}), (\ref{17.6}), and (\ref{24.1}). To
simplify these equations, we express $\gamma$ from (\ref{17.5}) as
\begin{equation}
    \gamma=\dsfrac{4y^3(2+\tilde{\alpha}^2)-x^2(3\pi-4x)}{6\tilde{\alpha}^2\pi^2} \; ,
    \label{24.1'}
\end{equation}
and then inserting it to (\ref{17.6}), we obtain
\begin{equation}
    8(x^3+2y^3)(1+\tilde{\alpha}^2)-3\pi\tilde{\alpha}^2(x^2+y^2)-6x^2\pi=0 \; ,
    \label{24.2}
\end{equation}
where $x\equiv s_d\ge0$ and $y\equiv s_s\ge0$ are yet unknown.
Now using the explicit expressions for the densities, given by Eqs. (\ref{17.1}) - (\ref{17.2})
we present (\ref{24.1}) as
\begin{equation}
   \ba
(x^3-y^3)^2-\dsfrac{2(3x^6+8x^3y^3+3y^6)}{\tilde{\alpha}^2}+\dsfrac{12\pi(x^5+y^5)}{5\tilde{\alpha}^2}
-\dsfrac{9\pi^4\gamma^2}{2}=0 ; .
\ea
    \label{24.3}
\end{equation}
For an arbitrary positive $\tilde{\alpha}^2$ and $\gamma$, given by Eq.  \re{24.1'},
the equations (\ref{24.2}) and (\ref{24.3}) can be solved numerically with respect to $x$
and $y$. Then, using these solutions in (\ref{24.1'}), the corresponding $\gamma$ can be
determined. As a result, we obtain the phase diagram presented in Fig. 6a. Here the solid
line separates the stable and unstable states, while the dashed line is the boundary between
the droplet and gaseous phases. For comparison, in Fig. 6b we present the similar phase diagram
of the Petrov's model. It is seen that the latter model leads to a rather large area for the
emergence of a droplet state, since, it neglects the unstable region with $c_d^2<0$. Note that
in this  case
\begin{equation}
   \tilde\cale_{tot}(Petrov)=\dsfrac{2\gamma (256\sqrt{\pi\gamma}-15\pi\tilde{\alpha}^2)}{15}=\tilde\cale_{GS}|_{\rho_0\approx\rho}+\tilde\cale_{LHY}|_{c_d=0}
\end{equation}
and, hence, the boundary, shown in Fig. 6b may be described by the relation
\begin{equation}
	256\sqrt{\pi\gamma}=15\pi\tilde{\alpha}^2=15\pi\dsfrac{|\delta g|}{g} \; .
	\lab{petrrelat}
\end{equation}
%%%%%%%%%%  FIG6 
%\newpage
\begin{figure}[H]
	\begin{minipage}[H]{0.49\linewidth}
		\center{\includegraphics[width=1.1\linewidth]{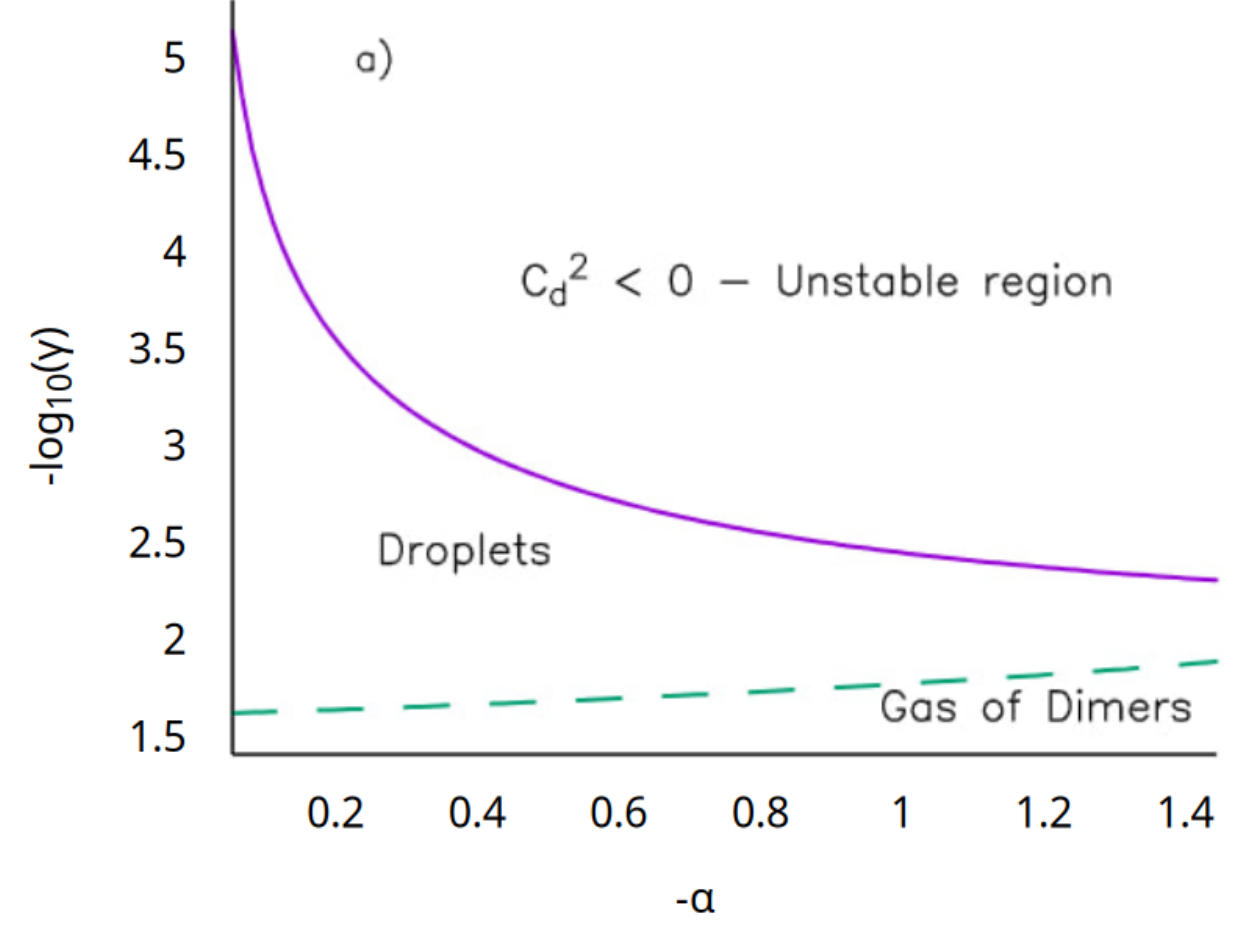} }
	\end{minipage}
	\hfill
	\begin{minipage}[H]{0.49\linewidth}
		\center{\includegraphics[width=1.1\linewidth]{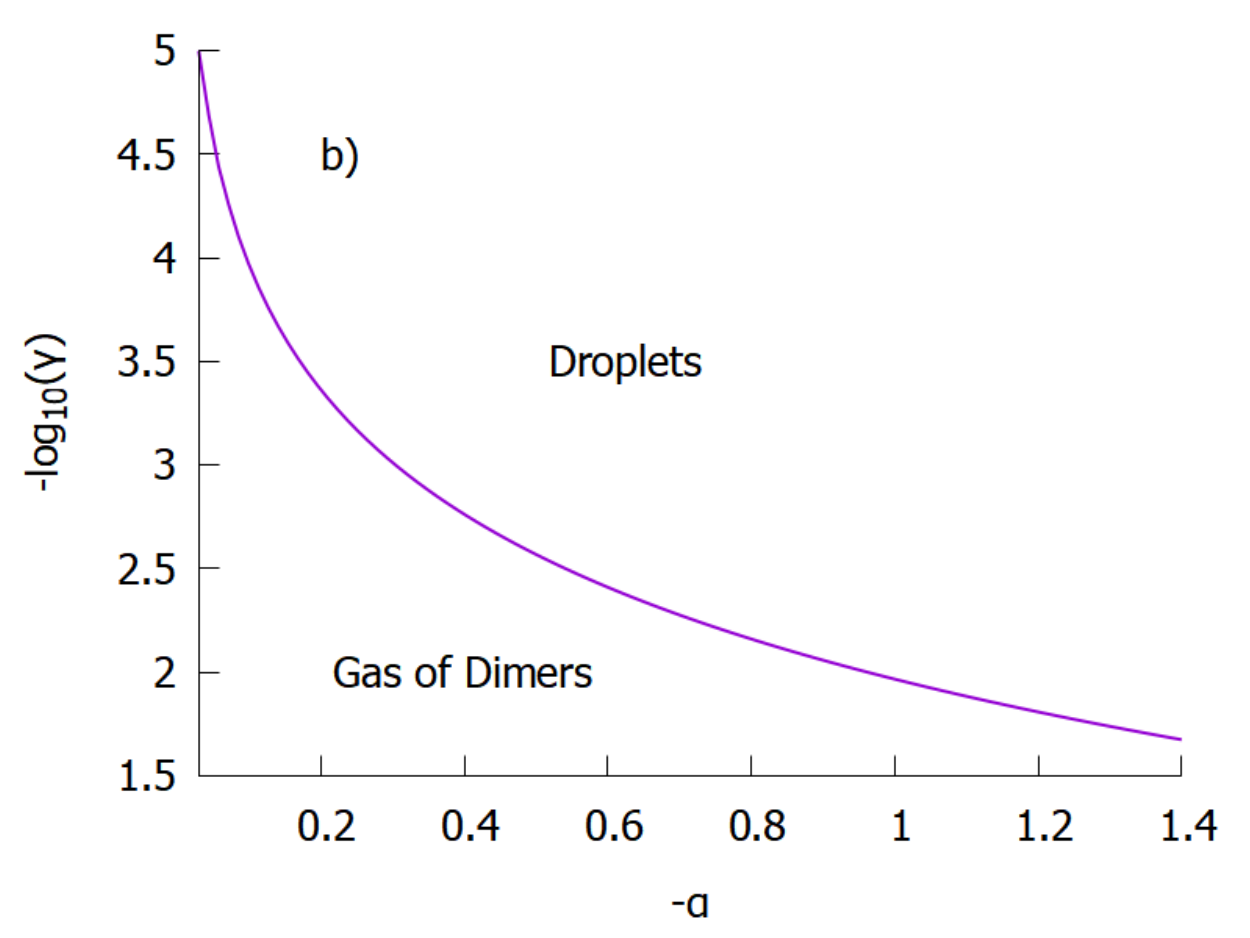} }
	\end{minipage}
	\caption{ The phase diagram of a symmetric binary Bose mixture at $T=0$ in the present 
		theory (a) and in the Petrov's model(b). }
	\label{Fig6}
\end{figure}

\section{Conclusion}
The Petrov's model, as well as the bilinear approximation, lead to the dynamical instability
of a binary Bose mixture
with attractive inter-species interactions $(\delta g/g<0)$. The instability is caused by
the fact that one of the branches of the energy dispersion, and hence, the related sound
velocity, becomes complex, $c_d^2<0$. In the present paper, we show that the solution of
the stability problem consists in taking account of all types of fluctuations characterized
by the intra-component normal as well as the symmetry-broken anomalous averages ($\sigma$),
and also including the mixed densities $(\rho_{ab})$. This procedure can be performed, and
the problem of $c_d^2\leq0$ solved self-consistently, in the framework of the optimized
perturbation theory, which is equivalent to the generalization of the Hartree-Fock-Bogoliubov
approximation from the single-component case to the case of multicomponent mixtures.

Detailed analysis, including numerical calculations, is accomplished for the unpolarized
symmetric binary Bose mixture at zero temperature. We find the boundary on the phase diagram
of the $(\gamma,-\delta g/g)$ plane, which separates the stable and unstable regions. As
expected, the stability region vanishes when $\sigma$ and $\rho_{ab}$ are omitted. Within
the region of stability, i.e. when the parameters $\gamma$ and $\delta g/g$ are appropriately
tuned, it is favorable for the system to be in the liquid droplet phase or in the phase of
a gas of dimerized atoms. We have found that, the condition for the emergency of the liquid
phase is rather subtle.

Unfortunately, experimental data for the energy dispersion of Bose mixtures, in the regime of
$(\delta g<0)$, to our knowledge, are not available. Comparing the present results with those
of Quantum Monte Carlo (QMC) calculations \ci{chiko2018}, it is possible to conclude that
they are in good qualitative agreement, although a more detailed comparison requires to take
into account finite-range corrections \ci{chiko2020}.

We are aware that, the droplet formation is preferable in polarized systems under the condition
$\rho_a/\rho_b=\sqrt{g_a/g_b}\ne1$ \ci{semegini}. This case, and the corresponding density
profiles of the self-bound droplets, within the presented theory, will be the subject of our
following studies.

\section*{Acknowledgments}
%\ack

The work is partly financed by the state budget of the Republic of Uzbekistan.
We are grateful to M. Nishonov for his assistance in preparation of the manuscript.

\vskip 3mm
\newpage
%\appendix
\section*{Appendix A} 
%{\bf A}
%\subsection{A}
\def\theequation{A.\arabic{equation}}
\setcounter{equation}{0}
%\flushright{
	%{\Large\bf Appendix 1}
	%}
%\bc {\Large\bf Appendix A} \ec \indent

In order to understand how the problem of complex sound velocity arises, let us outline the
basic points of the so-called bilinear (or Gaussian) approximation that is widely used for
condensed Bose systems.

After the Bogolubov shift \re{10.4} the action \re{10.3} is approximated by
\begin{equation}
	%   \begin{array}{cc}
		\ba
		S\approx S_0+S_2\;,\\
		\\
		S_0=\beta V\{-\mu_a\rho_{0a}-\mu_b\rho_{0b}+\dsfrac{1}{2}(g_a\rho_{0a}^2+g_b\rho_{0b}^2+2g_{ab}\rho_{0a}\rho_{0b})\} \; ,\\
		\\
		S_2=\dsint d\tau d\mathbf{r} \{\tilde{\psi}^\dagger \hat{K}_a\tilde{\psi}+\tilde{\phi}^\dagger \hat{K}_b\tilde{\phi}
		+ \dsfrac{g_{a} \rho_{0 a}}{2}
		[\tilde{\psi}^{\dagger2}+\tilde{\psi}^2+4\tilde{\psi}^\dagger \tilde{\psi}]+\\
		\\
		\dsfrac{g_{b}\rho_{0b}}{2}[\tilde{\phi}^{\dagger2}+\tilde{\phi}^2+4\tilde{\phi}^\dagger \tilde{\phi}] +\\
		\\
		g_{ab} [\rho_{0a}\tilde{\phi}^+\tilde{\phi}+
		\rho_{0b}\tilde{\psi}^+\tilde{\psi}+\sqrt{\rho_{0a}\rho_{0b}}
		(\tilde{\phi}\tilde{\psi}+\tilde{\psi}\tilde{\phi}^\dagger+h.c.)]\} \; .
	\end{array}
	\label{a10.5}
\end{equation}
where only quadratic terms in $\tilde{\psi}$ and $\tilde{\phi} $
are taken into account.
In the Cartesian representation
\begin{equation}
	\begin{array}{cc}
		& \tilde{\psi}=\dsfrac{1}{\sqrt{2}}(\psi_1+i\psi_2),   \  \  \tilde{\phi}=\dsfrac{1}{\sqrt{2}}(\psi_3+i\psi_4) \; ,
		
	\end{array}
	\label{a11.1}
\end{equation}
the quadratic term in (\ref{a10.5}) has the form:
\be
\ba
S_2 = \dsfrac{1}{2} \int d\tau \, d\mathbf{r} \left\{
\sum_{i=1}^4 \psi_i \left( -\dsfrac{\mathbf{\nabla}^2}{2m_i} + X_i \right) \psi_i
+ i \dssum_{i,j=1,2} \psi_i \, \partial_{\tau} \psi_j \, \varepsilon_{ij} \right. +\\
\left.
i\dssum_{m,n=3,4} \psi_m \, \partial_{\tau} \psi_n \, \varepsilon_{mn}
+ 2g_{ab} \sqrt{\rho_{0a} \rho_{0b}} (\psi_1 \psi_3 + \psi_3 \psi_1)
\right\} \; ,
\ea
\label{aS2momu}
\ee
where $\varepsilon_{ij}$ is an antisymmetric tensor, $m_{1,2}=m_a$, $m_{3,4}=m_b$, $(i=3,4)$, and
\begin{equation}
	\begin{array}{cc}
		&  X_1=-\mu_a+3g_a\rho_{0a}+g_{ab}\rho_{0b} \; ,\\
		& X_2=-\mu_a+g_a\rho_{0a}+g_{ab}\rho_{0b} \; ,\\
		& X_3=-\mu_b+3g_b\rho_{0b}+g_{ab}\rho_{0a} \; ,\\
		& X_4=-\mu_b+g_b\rho_{0b}+g_{ab}\rho_{0a} \; .
	\end{array}
	\label{a11.3}
\end{equation}
In the momentum space, we have
\begin{equation}
	\psi_i(\mathbf{r},\tau)=\dsfrac{1}{\sqrt{V\beta}}\sum_{n=-\infty}^{\infty}\sum_k\psi_i(\omega_n,\mathbf{k})e^{i\omega_n\tau+i\mathbf{kr}} \; ,
	\label{a11.4}
\end{equation}
where $\sum_k=V\int d\bf{k}/(2\pi)^3$ and $\omega_n=2\pi nT$ is the Matsubara frequency.
Equation  \re{aS2momu} can be represented as
\begin{equation}
	\begin{array}{cc}
		S_2=\dsfrac{(2\pi)^4}{2V\beta}\sum_{k,p,m,n}\sum_{i,j=1}^4\psi_i(\omega_n,\mathbf{k}) D_{ij}^{-1}(\omega_n,\mathbf{k},\omega_m,\mathbf{p})\times\\
		\psi_j(\omega_m,\mathbf{p})
		\delta(\mathbf{k}+\mathbf{p})\delta(\omega_m+\omega_n) \; ,
		\label{a11.5}
	\end{array}
\end{equation}
with the inverse propagator

\begin{equation}
	D^{-1}(\omega_n,\bfk)=
\begin{pmatrix}
		\veps_a(k)+X_1 & \omega_n & X_5 &0 \cr
		-\omega_n    & \veps_a(k)+X_2 & 0 & 0 \cr
		X_5 & 0 &\veps_b(k)+X_3 &\omega_n \cr
		0 & 0 &-\omega_n & \veps_b(k)+X_4 \cr
		\end{pmatrix}
\label{a12.1}
\end{equation}
where $\varepsilon_{a,b}(\mathbf{k})={\mathbf{k}}^2/{2m_{a,b}} $ and
\begin{equation}
	X_5=2g_{ab}\sqrt{\rho_{0b}\rho_{0b}} \; .
	\label{a12.111}
\end{equation}

The determinant of the inverse propagator gives rise to the following branches of energy
dispersion: $\omega_d\equiv \sqrt{\omega_{+}^{2}}$, $ \omega_s \equiv\sqrt{\omega_{-}^{2}} $ \; ,
\begin{equation}
	\begin{array}{cc}
		&  \omega_{\pm}^2=\dsfrac{E_a^2+E_b^2}{2}\pm \dsfrac{\sqrt{D_s}}{2} \; , \\
		& D_s=(E_a^2-E_b^2)+4E_{ab}^2 X_5^2 \; , \\
		& E_a^2=(\varepsilon_a(\bfk)+X_1)(\varepsilon_a(\bfk)+X_2) \; , \\
		& E_b^2=(\varepsilon_b(\bfk)+X_3)(\varepsilon_b(\bfk)+X_4) \; , \\
		& E_{ab}^2=(\varepsilon_a(\bfk)+X_2)(\varepsilon_b(\bfk)+X_4) \; .
	\end{array}
	\label{a12.2}
\end{equation}
Now, keeping in mind that both components of the mixture are Bose-condensed, we can exploit the
Hugenholtz - Pine relations \cite{HP,Watabe,Nepom}:
\begin{equation}
	\begin{array}{cc}
		&  \sum_n^a-\sum_{an}^a-\mu_a=0 \; , \\
		& \sum_n^b-\sum_{an}^b-\mu_b=0 \; ,
		\label{a13shtrix1}
	\end{array}
\end{equation}
where $\sum_n$ and $\sum_{an}$ are the normal and anomalous self -- energies, respectively.
In our notation, the relations (\ref{a13shtrix1}) have the simple form
% \cite{andersen}:
\begin{equation}
	X_2=X_4=0
	\label{a13shtrix2} \; .
\end{equation}
With this requirement, Eqs. (\ref{a11.3}) and (\ref{a12.111}) can be rewritten as
\begin{equation}
	\begin{array}{cc}
		&  \Delta_a=\dsfrac{X_1}{2}=g_a\rho_{0a} \; , \\
		\\
		&  \Delta_b=\dsfrac{X_2}{2}=g_b\rho_{0b} \; , \\
		\\
		&  \Delta_{ab}=\dsfrac{X_5}{2}=g_{ab}\sqrt{\rho_{0a}\rho_{0b}} \; .
	\end{array}
	\label{a13shtrix3}
\end{equation}
Moreover, for the equal masses, $m_a=m_b=m$ the dispersions in Eqs. \re{a12.2} are simplified:
\begin{equation}
	\begin{array}{cc}
		&  \omega_{\pm}^2(k)=\varepsilon_k^2+\varepsilon_k(\Delta_a+\Delta_b\pm \sqrt{\tilde{D}}) \; , \\
		& \tilde{D}=(\Delta_a-\Delta_b)^2+4\Delta_{ab}^2 \; ,
	\end{array}
	\label{a13shtrix4}
\end{equation}
with $\varepsilon_k=\mathbf{k}^2/{2m}$. The partition function $Z$ and the thermodynamic
potential can be evaluated directly from Eqs. (\ref{10.2}), (\ref{10.3}) and (\ref{a10.5})
by using the standart formula
%\begin{equation}
%    \begin{array}{cc}
	\be
	\ba
	
	Z(j_1,j_2)=\dsint \prod_{a=1}^{4} D\psi_ae^{-\frac{1}{2}\int dxdx' \psi_a(x)D_{ab}^{-1}(x,x')\psi_b(x')}\times \\
	e^{\int dx j_a(x)\psi_a(x)}=
	(Det D^{-1})^{(-1/2)}e^{\frac{1}{2}\int dxdx'j_a(x)D_{ab}(x,x')j_b(x')}.
	% \exp[\frac{1}{2}\int dxdx'j_a(x)D_{ab}(x,x')j_b(x')] \; .
	\label{a12.3}
	\ea
	\ee
	%    \end{array}
%
%\end{equation}
As a result, for $T=0$ one obtains
\begin{equation}
	\begin{array}{cc}
		& \Omega =\Omega_0+\Omega_{LHY} \; , \\
		& \Omega_0=V[-\mu_a\rho_{0a}-\mu_b\rho_{0b}+\dsfrac{g_a\rho_{0a}^2}{2}+
		\dsfrac{g_b\rho_{0b}^2}{2}+g_{ab}\rho_{0a}\rho_{0b}] \; , \\
		& \Omega_{LHY}=\dsfrac{1}{2}\dssum_k[\omega_+(k)+\omega_-(k)+counter\;  terms] \; .
	\end{array}
	\label{a12.4}
\end{equation}
It is seen that the bilinear approximation directly leads to the LHY term, which does not need
to be included artificially.

The question arises, whether this approach faces with the problem of complex sound velocity
for $g_{ab}<0$. To answer this question, it is sufficient to consider the symmetric balanced
case.

\vskip 2mm

{\bf Sound velocity in Bilinear approximation }

\vskip 2mm

Let $m_a=m_b=m$, $g_a=g_b=g$, $\mu_a=\mu_b=\mu$, $\rho_a=\rho_b=\rho/2$, and hence
$\rho_{0a}=\rho_{0b}=\rho_0/2$, $\Delta_a=\Delta_b=g\rho_{0a}$. Now, from (\ref{a13shtrix3})
and (\ref{a13shtrix4}), we obtain
\begin{equation}
	\omega_{\pm}(k)=\sqrt{\varepsilon_k[\varepsilon_k+2\rho_{0a}(g\pm g_{ab})]}
	\approx c_{\pm}k+O(k^3) \; ,
	\label{a13.1}
\end{equation}
which leads to following sound velocities:
\begin{equation}
	c_d^2\equiv c_+^2=\dsfrac{\rho_{0a}(g+g_{ab})}{m}=\dsfrac{\delta g \rho_{0a}}{m}\leq0 \; ,
	\label{a13.2}
\end{equation}
\begin{equation}
	c_s^2\equiv   c_-^2=\dsfrac{\rho_{0a}(g-g_{ab})}{m}=\dsfrac{ \rho_{0a}(2g-\delta g)}{m}\ge0 \; .
	\label{a13.3}
\end{equation}
Note that, in the approximation, where $\rho_{0a} \approx \rho_a$, we come back to the
Petrov's result (\ref{Cd Cs petr}). In (\ref{a13.2}) and (\ref{a13.3}), the condensed fraction
$\rho_{0a}$ and the inter-species coupling constant $g$ are positive, but $\delta g$ is
negative under the accepted condition. Therefore we see that, the bilinear approximation,
which takes into account only quadratic fluctuations by neglecting the anomalous density, is
not able to improve the situation related to the complex sound velocity.

%\subsection{B}
\section*{Appendix B}
\def\theequation{B.\arabic{equation}}
\setcounter{equation}{0}
%\bc {\Large\bf Appendix B} \ec \indent
Here we recall some principal points of the approach we use. For the simplicity of presenting
the main ideas, we consider the case of a single-component equilibrium uniform Bose
system. The occurrence of Bose-Einstein condensation implies global gauge symmetry breaking
\cite{Yukalov_2005,Yukalov_2025,yukalovannals,Lieb_2005}, which can be effectively
done by means of the Bogolubov shift \cite{Bogolubov_1967,Bogolubov_1970} that introduces
two variables,
\be
\label{A1}
 \psi(\bfr) = \eta + \widetilde\psi(\bfr) \;  ,
\ee
where $\eta$ is the Bose-condensate amplitude and $\tilde{\psi}$ is a field operator of
non-condensed particles. Here $\eta = \sqrt{\rho_0}$, while $\rho_0$ is the condensate
density, normalized to the number of condensed particles,
\be
\label{A2}
N_0 = N \; |\;\eta\; |^2 \qquad ( N_0 > 0 ) \; ,
\ee
with $N$ being the total number of particles. The number of condensed particles $N_0$, by
the Bogolubov-Ginibre theorem \cite{Bogolubov_1967,Bogolubov_1970,Ginibre_1968}, has to be
a minimizer of the thermodynamic potential, thus guaranteeing the system stability. Then
for the free energy $F$ we have
\be
\label{A3}
 \frac{\partial F}{\partial N_0} = \left\langle
\frac{\partial\hat H}{\partial N_0} \right\rangle = 0  \qquad
( N_0 > 0 ) \; ,
\ee
where $\hat{H}$ is the energy Hamiltonian. In that way, the Bogolubov-Ginibre conditions
(\ref{A2}) and (\ref{A3}) prescribe the value for the number of condensed particles $N_0$
and guarantee the stability of a system with broken gauge symmetry.

Since the total number of particles $N$ is assumed to be given and $N_0$ is fixed by the
Bogolubov-Ginibre stability condition, the average number of non-condensed particles
$N_1 = N - N_0$, defined as
\be
\label{A4}
N_1 = \langle \hat N_1  \rangle \; , \qquad
\hat N_1 = \int \widetilde\psi^\dagger(\bfr) \;
\widetilde\psi(\bfr) \; d\bfr \; ,
\ee
is also fixed. Under the broken global gauge symmetry, the number-conserving condition
(\ref{A4}) replaces the conservation condition for the total number $N = N_0 + N_1$ of
particles.

According to the basic principles of statistical physics, the system is correctly defined
provided it is characterized by a representative ensemble taking into account all conditions
uniquely defining the system \cite{Gibbs_1928,Gibbs_1931,Ter_1954,Jaynes_1957,Jaynes}. The
related statistical operator $\hat{\rho}$ is defined by the minimization of the entropy
$S = -{\rm Tr} \hat{\rho} \ln \hat{\rho}$ under the given constraints. In the case of the
system with broken gauge symmetry, in addition to the standard conditions of the statistical
operator normalization and energy definition,
\be
\label{A5}
 {\rm Tr} \hat{\rho} = 1 \; , \qquad
E = \langle \hat H \rangle =  {\rm Tr} \hat{\rho} \hat H \;  ,
\ee
we have the Bogolubov-Ginibre condition (\ref{A3}) and normalization condition (\ref{A4}).
Thus, we need to minimize the information functional
$$
I[\; \hat\rho\; ] = - S + \lambda_0 ( {\rm Tr}\hat\rho - 1) +
\beta ( {\rm Tr}\hat\rho \hat H - E ) +
$$
\be
\label{A6}
+
\beta \mu_0 ( N_0 - {\rm Tr}\hat\rho N_0 ) +
\beta \mu_1 ( N_1 - {\rm Tr}\hat\rho \hat N_1 ) \;  ,
\ee
in which $\lambda_0$, $\beta = 1/T$, $\mu_0$ and $\mu_1$ are the Lagrange multipliers.

Minimizing this information functional yields the statistical operator
\be
\label{A7}
 \hat\rho = \frac{1}{Z} \; e^{-\beta H} \qquad
\left( Z \equiv {\rm Tr} e^{-\beta H} \right) \;  ,
\ee
with the grand Hamiltonian
\be
\label{A8}
 H = \hat H - \mu_0 N_0 - \mu_1 \hat N_1 \;  .
\ee

The system chemical potential $\mu$, corresponding to the conservation of the total number
of particles $N$, is defined in the standard way yielding the expression
\be
\label{A9}
 \mu =\frac{\partial F}{\partial N} =
\frac{\partial F}{\partial N_0} \; \frac{\partial N_0}{\partial N} +
\frac{\partial F}{\partial N_1} \; \frac{\partial N_1}{\partial N} =
\mu_0 n_0 + \mu_1 n_1 \; ,
\ee
where the fractions of condensed and non-condensed particles, respectively, are
\be
\label{A10}
n_0  \equiv \frac{N_0}{N} \; , \qquad
n_1  \equiv \frac{N_1}{N} \;   .
\ee

The use of the representative ensemble, with two chemical potentials, makes the theory
self-consistent. Otherwise, as has been emphasized by Hohenberg and Martin \cite{HohenbergM},
the theory confronts the dilemma of either getting an unphysical gap in the spectrum or
violating thermodynamic relations, emplying the system instability. Of course, in a normal,
not Bose-condensed, system, where there is no gauge symmetry breaking, there is a need of
only a single chemical potential. In a Bose-condensed system with broken gauge symmetry,
it is possible to use a single chemical potential only under asyptotically weak interactions
and close-to-zero temperature, when almost all the system is Bose-condensed so that the
Bogolubov (bilinear) approximation is valid. But as soon as the condensate density essentially
deviates from the average particle density, it is necessary to resort to two chemical
potentials in order to avoid the Hohenberg-Martin dilemma. The theory becomes self-consistent
for the arbitrary strength of interactions only when both, Bogolubov-Ginibre stability condition
(\ref{A2}) and (\ref{A3}), as well as the normalization condition (\ref{A4}), are valid. Two
conditions require two chemical potentials playing the role of Lagrange multipliers.

For a nonunform system, $\eta$ becomes a condensate function $\eta({\bf r})$ and the stability
condition (\ref{A3}) takes the form of the variational derivative
 \be
\label{A11}
\frac{\delta F}{\delta\eta(\bfr)} = \left\langle
\frac{\delta\hat H}{\delta\eta(\bfr)} \right\rangle \; .
\ee
This approach is straightforwardly generalized to nonequilibrium systems, where, instead
of the extremization of the thermodynamic potential, the extremization of an action functional
is accomplished. All details can be found in Refs.
\cite{Yukalov_2005,Yukalov_2025,yukalovannals}.

%\subsection{C}
\section*{Appendix C}
\def\theequation{C.\arabic{equation}}
\setcounter{equation}{0}
%\bc {\Large\bf Appendix C} \ec \indent
The accuracy of the approach has been tested by accomplishing numerical calculations for
single-component systems with the Bose-Einstein condensate. For weak interactions, or low density,
we return back to the gaseous approximation. Thus, for a uniform system under weak interactions
at zero temperature, we obtain the Lee-Huang-Yang form \cite{Lee_1,Lee_2,Lee_3} of the energy
\be
\label{B1}
 E_N = \frac{\rho f_0}{2m} \; \left( 1 +
\frac{16}{15\pi^2} \; \sqrt{\rho f_0^3} \right) \; N \;  ,
\ee
where $\rho$ is the average density and $f_0 = 4 \pi a_s$ is a scattering amplitude.
From here, the sound velocity immediately follows, defined by the formula
\be
\label{B2}
 c^2 = \frac{1}{m} \; \left( \frac{\partial P}{\partial\rho} \right)_{TN} \;  ,
\ee
with the pressure
\be
\label{B3}
 P = - \left( \frac{\partial E_N}{\partial V} \right)_N =
\frac{\rho^2}{N} \; \left( \frac{\partial E_N}{\partial \rho} \right)_N  ,
\ee
which gives
\be
\label{B4}
 P = \frac{\rho^2 f_0}{2m} \; \left( 1 +
\frac{24}{15\pi^2} \; \sqrt{\rho f_0^3} \right)
\ee
and, respectively, the sound velocity
\be
\label{B5}
 c = \frac{\sqrt{\rho f_0}}{m} \; \left( 1 +
\frac{1}{\pi^2} \; \sqrt{\rho f_0^3} \right) \;  .
\ee

Notice that Beliaev \cite{belyaev}, using the second-order perturbation theory, found the
sound velocity
\be
\label{B6}
 c_B = \frac{\sqrt{\rho_0 f_0}}{m} \; \left( 1 +
\frac{7}{6\pi^2} \; \sqrt{\rho_0 f_0^3} \right) \;    ,
\ee
with the condensate density
\be
\label{B7}
 \rho_0 = \rho \; \left( 1 - \;
\frac{1}{3\pi^2} \; \sqrt{\rho f_0^3} \right) \;  .
\ee
Substituting (\ref{B7}) into (\ref{B6}) one gets the same expression (\ref{B5})
(hence $c_B = c$).

The most important is that our method allows for calculations at arbitrary interaction
strength. Thus, the ground-state energy in the whole range of stability is found
\cite{Yuk_Yuk_2014}, being in good agreement with Monte Carlo calculations \cite{Rossi}.
The condensate fraction for arbitraryly strong interactions is calculated \cite{Yuk_Yuk_2014},
being very close to Monte Carlo results \cite{Rossi}. The condensate fraction of trapped
atoms is derived \cite{Yukalov_JPB}, being in very good agreeement with Monte Carlo
simulations \cite{Du_1,Du_2}. The effect of trap-center condensate depletion caused by
strong interactions is explained \cite{Yukalov_JPB}. For the considered cases of
single-component systems the approach has demonstrated rather high accuracy.

\end{document}